\documentclass[final,5p,twocolumn]{elsarticle}
\usepackage[flushleft]{threeparttable}
\usepackage{graphicx}
\usepackage{makeidx}
\usepackage{amsmath}
\usepackage{array}
\usepackage{soul}

\usepackage{multirow}

\usepackage{amsfonts}%
\usepackage{amssymb}%
\usepackage[ruled,vlined]{algorithm2e}
\usepackage{etoolbox}
\makeatletter
\patchcmd{\algocf@Vline}{\vrule}{\vrule\hspace{-0.25em}}{}{}
\makeatother
\usepackage{graphics}
\usepackage{subfig}
\makeatletter
\g@addto@macro\endfrontmatter{\enlargethispage{-2\baselineskip}}
\makeatother

\usepackage[labelfont=bf]{caption}
\usepackage[font={small}]{caption}
\usepackage{array}
\usepackage{makecell}

\usepackage{transparent}
\usepackage{color}
\usepackage{xcolor}
\usepackage[utf8]{inputenc}
\usepackage[english]{babel}
\usepackage[hidelinks]{hyperref}

\makeatletter
\newcommand\footnoteref[1]{\protected@xdef\@thefnmark{\ref{#1}}\@footnotemark}
\makeatother
\usepackage{tabularx}
\usepackage{enumitem}
\usepackage{adjustbox}

\usepackage[figuresright]{rotating}

\setcitestyle{square}

\journal{NeuroImage}

\begin{document}
\begin{frontmatter}

\title{FastSurfer - A fast and accurate deep learning based neuroimaging pipeline}
\author[label1]{Leonie Henschel}
\author[label1]{Sailesh Conjeti}
\author[label1]{Santiago Estrada}
\author[label1]{Kersten Diers}
\author[label2,label3,label4]{Bruce Fischl}
\author[label1,label2,label3]{Martin Reuter\corref{cor1}}

\address[label1]{German Center for Neurodegenerative Diseases (DZNE), Bonn, Germany}
\address[label2]{A.A. Martinos Center for Biomedical Imaging, Massachusetts General Hospital, Boston MA, USA }
\address[label3]{Department of Radiology, Harvard Medical School, Boston MA,USA}
\address[label4]{Computer Science and Artificial Intelligence Laboratory, MIT, Cambridge MA, USA}

\cortext[cor1]{Data used in preparation of this article were obtained from the Alzheimer’s Disease Neuroimaging Initiative (ADNI)  database  (adni.loni.usc.edu).  As  such,  the  investigators  within  the  ADNI  contributed  to  the  design and implementation of ADNI and/or provided data but did not participate in analysis or writing of this report. A complete listing of ADNI investigators can be found at:http://adni.loni.usc.edu/wp-content/uploads/how\_to\_apply/ADNI\_Acknowledgement\_List.pdf}
\cortext[cor1]{Correspondence to: Martin Reuter (\texttt{martin.reuter@dzne.de}).}

\begin{abstract}
Traditional neuroimage analysis pipelines involve computationally intensive, time-consuming optimization steps, and thus, do not scale well to large cohort studies with thousands or tens of thousands of individuals. In this work we propose a fast and accurate deep learning based neuroimaging pipeline for the automated processing of structural human brain MRI scans, replicating FreeSurfer's anatomical segmentation including surface reconstruction and cortical parcellation. To this end, we introduce an advanced deep learning architecture capable of whole brain segmentation into 95 classes. The network architecture incorporates local and global competition via competitive dense blocks and competitive skip pathways, as well as multi-slice information aggregation that specifically tailor network performance towards accurate segmentation of both cortical and sub-cortical structures. Further, we perform fast cortical surface reconstruction and thickness analysis by introducing a spectral spherical embedding and by directly mapping the cortical labels from the image to the surface.
This approach provides a full FreeSurfer alternative for volumetric analysis (in under 1 minute) and surface-based thickness analysis (within only around 1h runtime). For sustainability of this approach we perform extensive validation: we assert high segmentation accuracy on several unseen datasets, measure generalizability and demonstrate increased test-retest reliability, and high sensitivity to group differences in dementia.
\end{abstract}
\end{frontmatter}

\begin{figure*}[h]
    \centering
    \includegraphics[width=\textwidth,keepaspectratio]{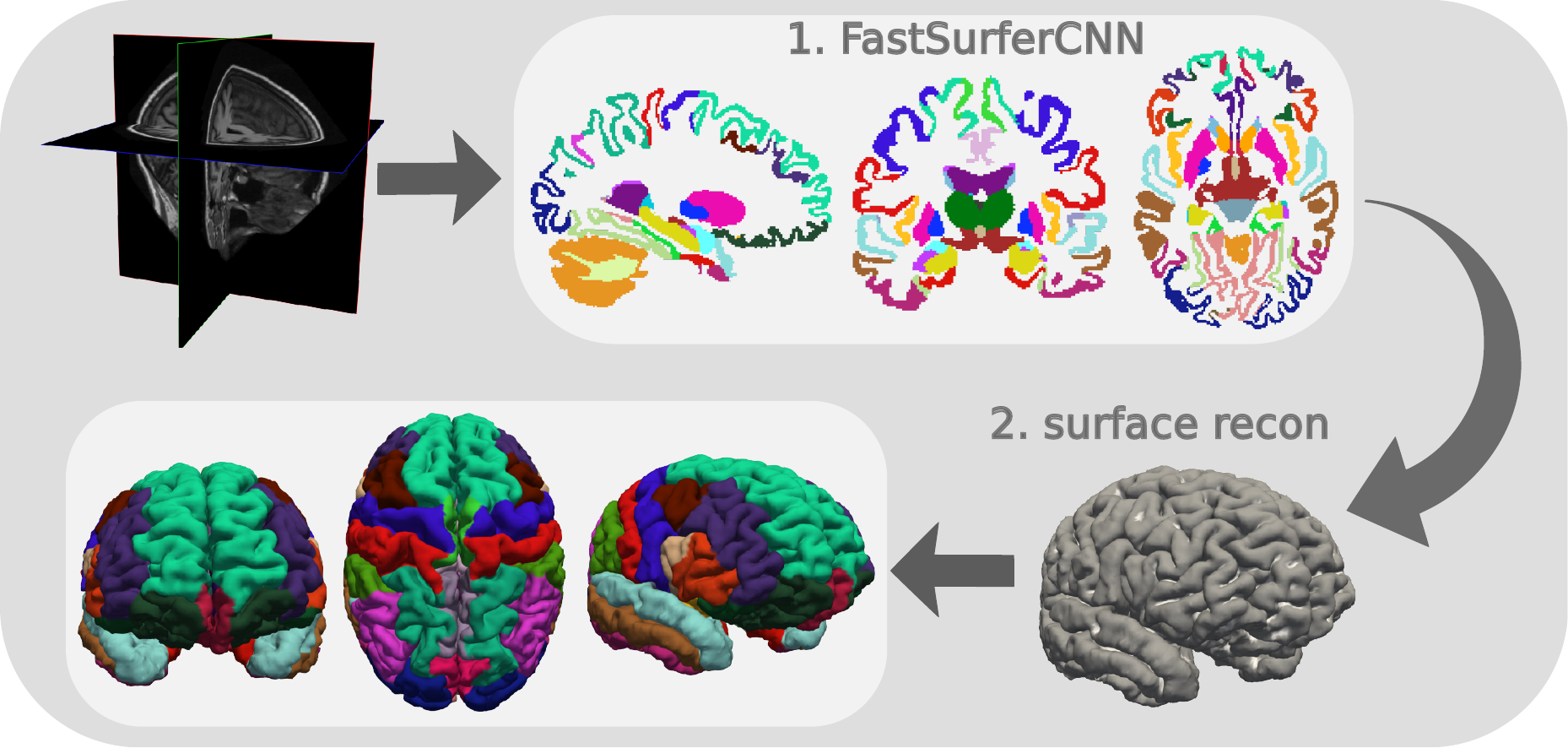}
    \caption{FastSurfer - A fast and accurate deep learning based neuroimaging pipeline}
    \label{fig:teaser}
\end{figure*}

\section{Introduction}

The rapid emergence of standardized robust non-invasive imaging methods and infrastructure for big data analysis over the years has promoted the advent of a variety of large-scale neuroimaging studies. Different initiatives aim to understand the variability, development and anatomical layout of the human brain in e.g.\  neurodegeneration (ADNI \cite{ADNI_dataset}, OASIS\cite{oasis_1_dataset, oasis_2_dataset}), psychiatric diseases (LA5c \cite{LA5c_dataset}), neurodevelopmental disorders (ABIDE \cite{abide_dataset}, MIRIAD \cite{miriad_dataset}) or within populations (Rotterdam Study \cite{rotterdam}, Human Connectome Project \cite{hcp_dataset}, UKBiobank \cite{ukbiobank}, Rhineland Study \cite{rhineland}). A core challenge within all neuroimaging studies is the need to process and analyze the continuing stream of data in a timely manner. As Magnetic Resonance Imaging (MRI) is one versatile imaging modality and integral part of all these studies, developing efficient tools to identify clinically-relevant imaging biomarkers with MRI are in high demand. In this work we, therefore, develop a fast method for volumetric segmentation, reconstruction of cortical geometry, and morphometric estimation of brain structures including cortical thickness. It is the first work that aims at integrating an advanced deep learning method for image segmentation into a complete processing pipeline that includes cortical surface reconstruction and  segmentation.

\subsection{Neuroimage Analysis}

To date, a few well maintained neuroimage processing pipelines such as FreeSurfer \cite{fischl2002whole}, BrainSuite \cite{Shattuck2002}, SPM \cite{spm_book2007}, ANTs \cite{Avants2009}, or FSL \cite{Jenkinson2012} are the only means available to process and evaluate the incoming flow of data. These pipelines usually employ multiple image transformation steps, some of which require careful fine-tuning of parameters such as convergence thresholds, smoothing levels, or iteration numbers. Furthermore, due to extensive numerical optimization, e.g.\ non-linear registration or Bayesian segmentation, these approaches are computationally expensive and suffer from long runtimes. Hence, several hours are required to process a single volume, significantly limiting scalability to large cohort studies with thousands of cases or to clinical workflows where immediate results are essential.

Supervised deep learning approaches are an attractive alternative to replace time-intensive steps within these pipelines such as whole-brain segmentation because of their faster runtimes. Fully convolutional neural networks (F-CNN), for example, are able to learn the correct feature representations in an end-to-end fashion from the image itself without requiring lengthy pre-processing steps. These methods can be effectively parallelized on graphical processing units (GPU) resulting in an enormous speed-up (seconds instead of hours). Additionally, these networks often outperform traditional approaches with respect to accuracy and have become increasingly popular for pixel or voxel-wise semantic segmentation tasks in computer vision and biomedical imaging~\cite{long2015fully,unet,noh2015learning,segnet,V-net,densenet}. In this work we propose a neuroimaging pipeline based on an advanced neural network architecture for whole brain segmentation that induces local and global competition in the dense block and skip-connections.  

\subsection{Deep Learning for Whole Brain Segmentation}
The task of whole-brain segmentation in particular is challenging due to the complex 3D architecture and spatial dependency between slices, the large number of labels, the size of the scanning volumes (memory requirements), and variability across scanners and subjects. While several deep learning based approaches have been proposed for specific tasks, such as tumor segmentation \cite{Rani2017, Dong2017, Arunachalam2017, Havaei2017, Amin2018, Crimi2019}, brain lesion segmentation \cite{Kamnistas2017, Varghese2016, Rezaei2017, RoaBarco2017, Chen2018}, MR image reconstruction \cite{Jin2017, Mardani2017, Schlemper2018, Yang2018, Dedmari2018}, prediction of brain related diseases and their progression \cite{Payan2015, Qi2016, HosseiniAsl2016, Lee_2019} or segmentation of a smaller number of brain (sub-)structures \cite{Zhang2015,Akkus2017, Milletari2017,Fedorov17,Dolz2018,Thyreau2018b,Chen2018VoxResNet,Nogovitsyn2019,Li2019,Sun2019,Ito2019} full brain segmentation into more than 25 classes has - so far - only been achieved by a few groups \cite{de_Brebisson_2015,Moeskops2016,Mehta2017,wachinger2018deepnat,sdnet,quicknat,psacnn,slant2019,assemblynet2019} - yet with the exception of \cite{quicknat} only with direct comparison of segmentation accuracy on a test-set, lacking extensive validations, e.g., of reliability and sensitivity to real neuroanatomical effects.

Most of these brain segmentation networks were trained on extracted 3D patches \cite{de_Brebisson_2015,wachinger2018deepnat,Mehta2017} or 2D slices \cite{Moeskops2016,sdnet,quicknat,psacnn}. Both approaches loose spatial information critical for correct classification of a given structure. Due to memory constrains of existing GPUs it is, however, currently not feasible to train a full 3D deep neural segmentation network with a large number of labels for a whole-brain MRI volume at full resolution. Even for a lower number of labels, batch size can usually not be larger than 1, which would be another detrimental factor. Segmenting downsampled images is also not an option, e.g.\ due to small cortical structures, that may not be sufficiently resolved at lower resolutions. 
Additionally, full 3D networks have a much higher demand for training data as one MRI volume constitutes a single training case.
Recently, approaches to overcome memory demands for whole-brain segmentation have been proposed, such as the combination of multiple (up to 250) patch-based 3D CNNs that process overlapping sub-volumes of the divided whole-brain input \cite{slant2019,assemblynet2019}. These segmentation tools, however, require a number of additional and potentially error-prone processing steps including image registration to a standard space and bias field removal, as well as inference for a large number of networks. Overall, most of the architectures proposed for brain segmentation so far have runtimes of around 15-60~min on the GPU \cite{de_Brebisson_2015,Moeskops2016,Mehta2017,wachinger2018deepnat,slant2019,assemblynet2019} which makes them unattractive for integration into a maximally fast neuroimaging pipeline. The networks presented in \cite{sdnet} and \cite{quicknat} can produce a whole brain segmentation in less than one minute and are thus most relevant to our work. 
 
The SkipDeconv-Net (SD-Net~\cite{sdnet}) was the first whole-brain segmentation F-CNN and is based on a classic encoder-decoder architecture reminiscent of the U-net~\cite{unet}. The SD-Net introduces a novel loss-function that addressed the inherent class imbalance problem and alleviated segmentation errors along anatomical boundaries. Subsequently, the network architecture was extended into an F-CNN called Quick segmentation of Neuroanatomy (QuickNAT~\cite{quicknat}), which allows segmentation of a whole 3D brain volume into 27 structures. In this architecture, short-range skip connections were employed within each encoder-decoder block - these dense blocks were introduced in ~\cite{denseconnections} for classification tasks. Further, QuickNAT uses three 2D F-CNNs to produce predictions for axial, coronal and sagittal slices which are combined in a final multi-view aggregation step and thus allows partial recapture of spatial information in the third dimension (e.g.\ 2.5D approach).
Here, we compare to the different 2D architectures (with multi-view aggregation switched on for all of them for better comparability) and additionally include a modified patch-based 3D version of the UNet \cite{3DUNet} which allows inference within a similar time frame. To this end, we use the self-adapting framework ”no-new-Net” (nnU-Net) \cite{nnUNet} to fine-tune a 3D UNet architecture to the given image geometry and number of classes in our application. 

Here, we propose FastSurferCNN a deep learning architecture capable of segmenting a whole brain into 95 classes in merely 1 minute on the GPU (and 14 minutes sequential processing on the CPU). The basic architecture is inspired by QuickNAT \cite{quicknat}. Each F-CNN has the same encoder/decoder-based architecture with skip connections \cite{unet}, enhanced with unpooling layers \cite{noh2015learning} and dense connections \cite{denseconnections} within each block. The main methodological improvements of FastSurferCNN compared to QuickNAT are the introduction of competition within each block (competitive dense blocks) by replacing concatenation with maxout operations \cite{maxout,Estrada2018,fatsegnet}, as well as the inclusion of a wider image context within each 2D F-CNN (spatial information aggregation) in order to retain enough spatial information for the accurate segmentation of neuroanatomical structures, such as cortical gray matter regions.  

Note, that voxel-based image segmentation, on its own, is limited with regard to neuroimage analysis and biomarker extraction. Especially surface-based analysis has proven pivotal for e.g.\ correct estimation of thickness - an issue which has so far not been addressed in comparative publications on deep learning.  Existing traditional pipelines go far beyond image segmentation and provide utilities such as creation of cortical surface models, estimation of thickness, construction of fiber tracts or functional connectivity graphs, and tools for group comparison, such as registration and statistical frameworks. A major focus of this work is to fill this gap by integrating the developed deep learning framework (FastSurferCNN) into a complete, self-contained imaging pipeline called FastSurfer. 

Starting from the accurate 3D whole brain segmentation, provided by our deep learning framework, we perform cortical surface reconstruction and fast spherical mapping via a novel spectral approach that quickly maps the cortex using Laplace Eigenfunctions. Furthermore, we map cortical labels and include traditional point-wise and ROI thickness analysis, resulting in a full FreeSurfer alternative with approximately 60~min runtime (depending on image quality and process parallelization) of which only 1~min is attributed to the whole brain segmentation. Hence, FastSurfer combines the speed of supervised deep learning approaches and the convenience of the broad spectrum of surfaced-based features and analysis methodologies provided by traditional neuroimaging pipelines.

We extensively validate the quality of our deep learning based neuroimaging pipeline through assessment of segmentation accuracy, generalizability to unseen datasets and acquisition parameters, test-retest reliability, and sensitivity to group level differences in imaging cohorts in a number of publicly available datasets. In fact, this is the first work within deep learning approaches with such an exhaustive validation. We demonstrate that despite being orders of magnitude \textit{faster} than traditional approaches, FastSurfer \textit{increases reliability and sensitivity} making it a dependable tool for future large-scale population analysis tasks. The source code of FastSurfer is available on Github: https://github.com/Deep-MI/FastSurfer.\footnote{Will be made public upon acceptance.}

\section{Methodology}

\subsection{Datasets}
MRI volumes from eight publicly available datasets were selected for training, testing, and for extensive validation of the FastSurfer pipeline. In brief, selected subjects from the Autism Brain Imaging Data Exchange II  (ABIDE II) \cite{abide_dataset}, the Alzheimer’s Disease Neuroimaging Initiative (ADNI) \cite{ADNI_dataset}, the UCLA Consortium for Neuropsychiatric Phenomics LA5c Study (LA5c) \cite{LA5c_dataset}, and  the Open Access Series of Imaging Studies 1 and 2 (OASIS-1 \cite{oasis_1_dataset} and OASIS-2 \cite{oasis_2_dataset}) were used for network training. Further, subjects never encountered during training from ADNI and OASIS-1, the Minimal Interval Resonance Imaging in Alzheimer’s Disease (MIRIAD) \cite{miriad_dataset}, the Human Connectome Project (HCP) \cite{hcp_dataset}, a multi-subject, multi-model neuroimaging dataset (MMND) \cite{MMND}, the Traveling Human Phantom (THP) \cite{HTP_dataset} as well as the Mindboggle-101 dataset \cite{Klein2012} were used for validation and testing of the FastSurfer pipeline. For a detailed description of these datasets as well as a usage summary see Appendix Section \ref{sec:datasets}. 

All datasets were processed using FreeSurfer v6.0. FreeSurfer is an open source neuroimage analysis suite  \cite{fischl2002whole,fischl2012freesurfer} (http://surfer.nmr.mgh.harvard.edu/). 
Freesurfer morphometric procedures have been demonstrated to show good test-retest reliability across scanner manufacturers and across field strengths \cite{Han2006,reuter:long12}. In this work FreeSurfer parcellation following the ``Desikan–Killiany–Tourville'' (DKT) protocol atlas \cite{Klein2012, Desikan2006} is used for training and evaluation. In order to limit the number of segmentation labels, cortical regions touching each other across the hemispheres, are lateralized while all others are combined thus reducing the total number of labels from 95 (DKT without corpus callosum segmentations which are added later) to 78 during network training. Association to the left or right hemisphere is restored in the final prediction by estimating the closest white matter centroid (left or right hemisphere) to each label cluster. A list of all segmentation labels is provided in the appendix (see Table~\ref{tab:labels}). In accordance with FreeSurfer, all MRI brain volumes are conformed to standard slice orientation and resolution (1~mm isotropic) before feeding them to the different deep learning networks. No further image processing is required afterwards (e.g.\ no skull stripping or intensity normalization). 

\begin{figure*}[ht]
    \centering
    \includegraphics[width=17cm,height=15cm,keepaspectratio]{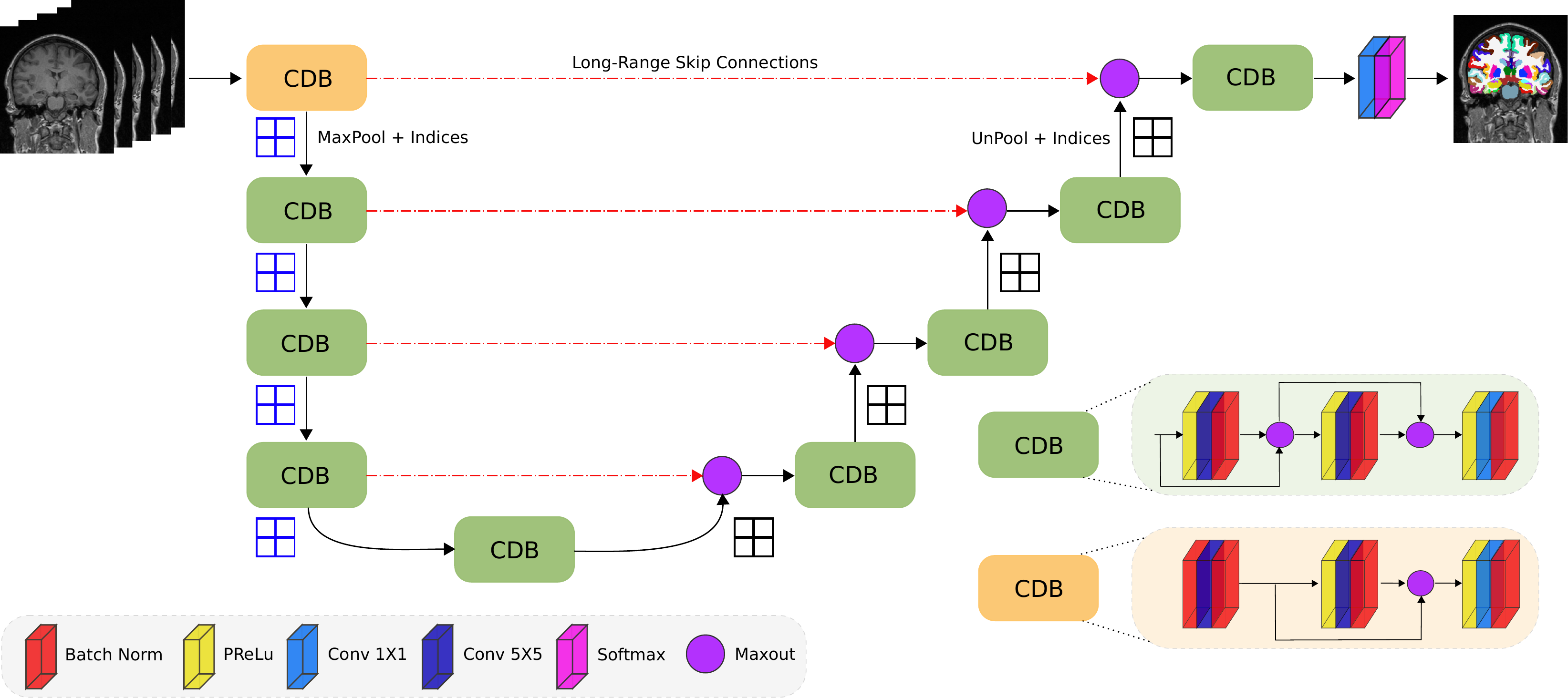}
    \caption{FastSurfer Network Architecture. The network consists of four competitive dense blocks (CDB) in the encoder and decoder part, separated by a bottleneck layer. Each block is composed of three sequences of parametric rectified linear unit (PReLU), convolution (Conv) and batch normalization (BN) with exception of the very first encoder block. In the first block, the PReLU is replaced with a BN to normalize the raw inputs.}
    \label{fig:CNN_architecture}
\end{figure*}

\subsection{FastSurfer CNN}
Here, we introduce the network architecture (FastSurferCNN) for whole brain segmentation into 95 classes (excluding background) in under 1 minute on the GPU (and approximately 14 minutes on the CPU). FastSurferCNN is composed of three F-CNNs operating on coronal, axial and sagittal 2D slices and a final view aggregation stage. The basic architecture of all three F-CNNs follows that of \cite{quicknat}, namely a sequence of 4 dense encoder and decoder blocks separated by a bottleneck layer as illustrated in Figure~\ref{fig:CNN_architecture}. Within FastSurferCNN, we integrate two improvements - competitive dense blocks and spatial information aggregation - targeted to promote information recovery and increase network connectivity. In the following sections, each of these elements will be explained in detail.

\begin{figure}[!hbt]
    \centering
    \includegraphics[width=8cm,height=8cm,keepaspectratio]{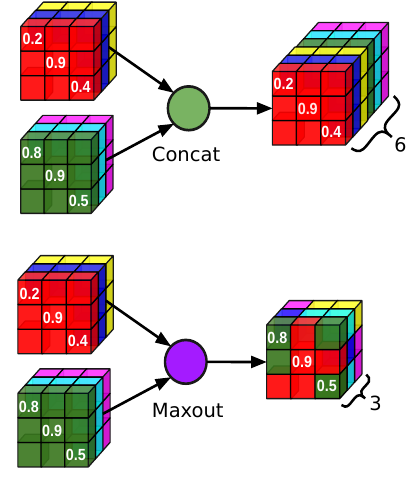}
    \caption{The standard ``concat'' operation (top) appends the two incoming blocks doubling memory requirements in each step. The efficient ``maxout'' operation (bottom) retains the max value at each position, hence inducing local competition between the blocks while keeping memory requirements constant.}
    \label{fig:maxout}
\end{figure}

\subsubsection{Competitive Dense Block}\label{sec:competition}
With the exception of our recent work in \cite{Estrada2018,fatsegnet}, dense connections within convolutional blocks have been implemented via concatenation of feature maps (see QuickNAT \cite{quicknat}) - effectively doubling the numbers of learnable parameters (Figure~\ref{fig:maxout} top) within each encoder and decoder block and thus considerably increasing memory requirements. Here, we employ competitive dense blocks in which concatenations are replaced with maxout activations \cite{maxout,Estrada2018,fatsegnet}. 
The maxout activations induce competition between feature maps and significantly reduce the number of parameters compared to the classical dense blocks, thus creating a lightweight model (Figure ~\ref{fig:maxout} bottom). Instead of stacking the output of previous layers on top of each other, only the maximum value at a given position is retained, keeping the number of input channels as well as parameters constant in each convolution layer. Assuming $L$ inputs, denoted as $\mathbf{X} = \left \{ \mathbf{x}^{l} \right \}_{l=1}^{L}  $, with each $\mathbf{x}^{l} = \left [ x_{ijk}^{l} \right ]_{i,j,k=1}^{H,W,C}$, where $H$ is height, $W$ is width and $C$ are number of channels for a particular feature map($\mathbf{x}^{l}$), the $\text{maxout}(\mathbf{X})$ output is given by: 
    
     \begin{equation}
     \begin{split}
         \text{maxout}(\mathbf{X}) = \left [ y_{ijk} \right]_{i,j,k=1}^{H,W,C} \\
         \text{ where }y_{ijk} = \text{max} \left \{ x^{1}_{ijk},\cdots,x^{L}_{ijk} \right \} .
     \end{split}
     \end{equation} 

The difference between dense blocks and competitive dense blocks can thus be described as:

\begin{eqnarray}
     \mathbf{X}_{l} = H_{3}^{l}(\mathbf{y}_{2}) \label{eq:2}\\ 
     \mathbf{y}_{2} = \text{concat}({H_{2}^{l}}(\mathbf{y}_{1}),\mathbf{y}_{1},\mathbf{X}_{l-1}) \\ 
     \underbrace{\mathbf{y}_{1} = \text{concat}({H_{1}^{l}}(\mathbf{X}_{l-1}),\mathbf{X}_{l-1})}_{
     \text{Densely Connected Block}}\label{eq:final_dense} \\ 
     \mathbf{X}_{l} = H_{3}^{l}(\mathbf{y}_{2})\label{eq:begin_max} \\ 
     \mathbf{y}_{2} = \text{maxout} (H_{2}^{l}(\mathbf{y}_{1}),\mathbf{y}_{1}) \\ 
     \underbrace{\mathbf{y}_{1} = \text{maxout} (H_{1}^{l}(\mathbf{X}_{l-1}),\mathbf{X}_{l-1})}_{\text{Competitive Dense Block}}\label{eq:densevsmaxout}
     \end{eqnarray}

Here, $H_{j}^{l}$ represents a composite function of three consecutive operations: parametric rectified linear unit (PReLU) followed by convolution and batch normalization (BN) with exception of the very first encoder block. In the first block, the raw inputs are passed through BN, convolution and another BN before following the previously described architecture (see Figure~\ref{fig:CNN_architecture}, CDBi vs.\ CDB block). The sequence of operations in $H_{j}^{l}$ guarantees normalized inputs to the maxout activation thereby improving convergence~\cite{normalisation} and increasing the exploratory span of the created sub-networks~\cite{competitive} simultaneously. Furthermore, filter co-adaptation is implicitly discouraged by the short-range skip-connections within the dense blocks~\cite{competitive}.

In addition to the competitive dense blocks we also implement competition across long-range skip connections. Instead of concatenating the unpooled information from the decoder arm with the corresponding feature maps from the encoder arm, we perform a maxout operation before further feeding the inputs to the competitive dense decoder blocks. All our competitive dense blocks are designed such that the inputs to this operation are already normalized (unpooling and skip transfer after BN; see Figure \ref{fig:CNN_architecture}). 

\subsubsection{Spatial information aggregation}\label{sec:spatial_info}

As mentioned in the introduction, training a full 3D deep neural network for whole-brain segmentation is currently not feasible for a large number of classes. However, 2D networks with single slice inputs loose information on the 3D spatial dependency between the inputs which can be crucial for correct segmentation of neuroanatomical structures. In order to retain as much information as possible, we provide the network with a larger volumetric context by passing a multi-slice input similar to \cite{Ding2017,Yu2018,Ghavami2018} where a sequence of neighbouring slices instead of a single 2D slice are used for network training.

Our spatial information aggregation approach consists of passing a 7-channel image by stacking the three preceding, the current, and the three succeeding slices for segmenting only the middle slice. Fundamentally, this spatial information aggregation combines the advantages of 3D patches (local neighbourhood) and 2D slices (global view). We, furthermore, analyze the impact of multi-slice images by comparing directly to single slice inputs.

\subsubsection{View Aggregation}\label{sec:view_agg}
In order to account for the inherent 3D geometry of the brain, one F-CNN per anatomical plane is trained and their outputs combined in a final view aggregation step. Depending on the orientation of the 2D slices, each network therefore learns the anatomical representation of the brain structures within the coronal, axial or sagittal view. The final segmentation is generated by aggregating the probability maps of each model through a weighted average. Combination of the three principal views can boost accuracy for cortical folds and subcortical structures, some of which are better represented in one of the individual planes. In addition, view aggregation acts as a regularizer to reduce erroneous predictions. As it is not possible to differentiate between the left and right hemispheres in the sagittal view, we merge lateral labels, effectively reducing the number of classes from 78 to 50 in the sagittal network. The probability maps of these lateralized classes are finally restored by copying the softmax output of the combined label to both left and right hemispheres. To account for this remapping step, the weight with which the sagittal predictions influence the final segmentation is reduced by one half compared to the other two views. 

\subsubsection{Model Learning}

\textbf{Training Dataset: } 140 representative subjects from ABIDE-II, ADNI, LA5C, and OASIS (see Section~\ref{sec:datasets}) were selected for training the F-CNN models and 20 subjects from MIRIAD were used for validation. Empty slices were filtered from the volumes, leaving on average 145 single view planes per subject and a total training size of above 20k images per network. In addition, we use data augmentation (random translation of maximally 16~mm) to artificially increase the training set size further.

The training set is balanced with regard to gender, age, diagnosis, and spans various other parameters (i.e.\ scanners, field strength, and acquisition parameters); the distribution of the subset is presented in Table~\ref{table:population}. Sufficient anatomical and acquisition variety in training images can be expected to improve network robustness, generalizability, and ultimately segmentation accuracy on most unseen scans without the need to fine-tune model weights. We will analyze generalizability to unseen datasets below. 

\begin{table}[!hbt]
\centering
\caption{Characteristics of the participants (n=160) showing mean ($\pm$ standard deviation) for continuous and counts (PCT) for categorical variables.}
\label{table:population}
\resizebox{\columnwidth}{!}{
\begin{threeparttable}
\begin{tabular}{l|cccc}
\textbf{Dataset} & \textbf{Subjects} & \textbf{Age ($\pm$ SD)}  &  \textbf{Women,n(\%)} & \textbf{Controls,n(\%)} \\ \hline
LA5c~\cite{LA5c_dataset} & 20 & 35 ($\pm$ 6.25) & 10 (50) & 10 (50) \\
MIRIAD~\cite{miriad_dataset}& 20 & 68.85 ($\pm$ 5.60)  & 10 (50) & 10 (50)\\
OASIS-1~\cite{oasis_1_dataset}& 40 & 37.48 ($\pm$ 13.54)  & 21 (52.5) & 40 (100) \\
OASIS-2~\cite{oasis_2_dataset}& 20 & 77.70 ($\pm$ 7.23)  & 11 (55) & 10 (50) \\ 
ABIDE II~\cite{abide_dataset} & 20 & 25.94 ($\pm$ 4.89)  & 10 (50) & 10 (50)\\ 
ADNI~\cite{ADNI_dataset} & 40 & 74.65 ($\pm$ 6.71)  & 20 (50) & 20 (50) \\ \hline
Total & 160 & 53.97($\pm$ 22.19) & 82 (51.25) & 100 (62.5) \\
\hline  

\end{tabular}
\end{threeparttable}}
\end{table}

\textbf{F-CNN Implementation: } Independent models for coronal, axial, and sagittal plane are implemented in PyTorch~\cite{Paszke2017} and trained for 30 epochs using two NVIDIA Titan Xp GPU with 12 GB RAM and the following parameters: batch size of 16, constant weight decay of 10$^{-04}$, and an initial learning rate of 0.01 decreased by 95~\% every 5 epochs. The networks are trained with Adam optimizer \cite{adamOptimizer} and a composite loss function of median frequency balanced logistic loss and Dice loss~\cite{sdnet}. This loss function encourages correct segmentation of tissue boundaries and counters class imbalances by up-weighting less frequent classes.

\subsection{FastSurfer Pipeline}\label{sec:pipeline}

Based on FreeSurfer methods and novel contributions, we also introduce a surface processing pipeline, that integrates our neural network architecture at its core to provide FreeSurfer volume and surface results, including cortical surfaces, thickness maps, and summary statistics in cortical regions following the DKT protocol atlas \cite{Klein2012, Desikan2006}.

Traditionally, surfaces are generated via a pipeline of several time-consuming processing steps: First, based on a white matter segmentation, which is patched to remove holes and ensure connectivity, initial surface triangle meshes are created for each hemisphere \cite{Dale1999, Fischl2001}. Meshes are smoothed, mapped to a sphere to localize topological defects (i.e.\ holes or handles should not exist as each hemisphere should be topologically equivalent to the sphere) \cite{Fischl1999a, Segonne2007}. Once all defects are fixed, surface placement along the white matter is fine-tuned and a second expanded surface (pial surface) is placed at the outer gray matter (GM) boundary, also providing thickness estimates at every point on the cortex \cite{Dale1999, Fischl2000}. Then, surfaces are carefully mapped to the sphere a second time (minimizing metric distortions), registered to a spherical atlas \cite{Fischl1999b}, and segmented into cortical parcellations (DKT atlas) \cite{Desikan2006, Fischl2004b, Klein2012}.

Based on the DKT volume segmentation available from the FastSurferCNN we modify the above FreeSurfer pipeline to yield surface results of FreeSurfer (including thickness and cortical ROI measures). A significant speed-up compared to FreeSurfer can be achieved by omitting several steps that have become obsolete, such as skull stripping and non-linear atlas registration, given that a high-quality full brain segmentation has already been achieved. Furthermore, we innovate some traditional approaches by developing novel modules based on spectral mesh processing. Specifically:
\begin{enumerate}
\item We use the full DKT brain volume segmentation to create a brainmask by closure, i.e.\ dilation and erosion, of the labels (including the ventricle label). This mask covers all labeled areas. Cortical regions are padded by one voxel layer to allow the pial surface to find its final position in some partial volume voxel between GM and CSF. Exceptions are the lateral orbital frontal and pars orbitalis to avoid capture of the optic nerve.
\item We retrospectively construct a quick bias field corrected brain image and a linear Talairach registration as these results are needed later for some relevant statistics (e.g.\ intracranial vault volume for head size estimation \cite{buckner:04}). Here we follow FreeSurfer, except that we can initialize the NU correct \cite{nu_correct} with the already existing brainmask.
\item We generate initial surfaces by using a marching cube \cite{Lorensen1987} algorithm rather than the traditional approach \cite{Dale1999} aiming at higher mesh quality at a slightly reduced number of vertices. 
\item We develop a fast mapping to the sphere using the Eigenfunctions of the Laplace operator to perform a spectral embedding of the original white matter surfaces quickly (for the topology fixer). Precisely, we solve the Laplace-Beltrami Eigenvalue problem $\Delta f = - \lambda f$ \cite{reuter:cad06,reuter:ijcv09} on the original cortical surface mesh to obtain the first three non-constant Eigenfunctions with smallest Eigenvalues. After correcting sign flips and swaps, these functions parametrize the surface smoothly in anterior-posterior, superior-inferior and lateral-medial directions. The spherical map can then be quickly obtained by projecting the 3D spectral embedding to the sphere, i.e.\ by scaling the 3D Eigenfunction vector to unit length for each vertex.

\item After topology fixing and GM surface creation, we map the DKT GM segmentations from the image onto the surface and compute surface ROI statistics, such as mean thickness and curvature averages per region - mimicking FreeSurfer's surface segmentation pipeline without requiring the non-linear spherical atlas registration and segmentation. Spherical atlas registration can, however, be included if cross-subject correspondence is required, e.g.\ for local surface-based thickness analysis.
\end{enumerate}

Overall, this yields a fast alternative to the FreeSurfer pipeline. We will evaluate the speed-up, reliability, and sensitivity of the full FastSurfer pipeline below.

\subsection{Statistics}

We will thoroughly validate the novel FastSurfer pipeline in terms of accuracy, generalizability, reliability and sensitivity using, Dice overlap, intraclass correlation and group analyses on volume and thickness ROI's, as well as thickness maps. In the following sections we explain these statistical methods in detail. 

\subsubsection{Dice similarity coefficient}
The Dice similarity coefficient (DSC) is a metric to evaluate the segmentation performance of the deep learning networks and can mathematically be expressed as 
\begin{align}
DSC(\textbf{G},\textbf{P}) = \frac{2 \times \textbf{G} \cap \textbf{P}}{\textbf{G} + \textbf{P}}
\label{eq:10}
\end{align}

with binary label maps of ground truth $\textbf{G}$ and prediction $\textbf{P}$ (pixels of the given class indicated with 1, all others with 0). Here, the DSC is used two fold: first, to directly compare the performance of different network architectures against each other, and second, to estimate similarity of the predictions achieved with FastSurferCNN and FreeSurfer v6.0 for a number of previously unseen datasets (generalizability). The DSC will be calculated separately for each cortical and sub-cortical structure. 
Note, that neural networks tend to smooth results, e.g.\, removing segmentation noise such as incorrect protrusions that are encountered in only a few training images and usually in random locations. While this kind of smoothing can improve segmentation accuracy, it can decrease the DSC which is partially affected by noise in the ground truth data. This is one reason why it is essential to perform additional validations (such as reliability or sensitivity analysis). 

\subsubsection{Average Hausdorff distance}
The average Hausdorff distance (AVG HD) is a metric for measuring the similarity between two sets of points and is often used to evaluate the quality of segmentation boundaries. It is defined as 
\begin{equation}
\begin{aligned}
AVG\ HD(\textbf{G}, \textbf{P}) ={} & \frac{1}{|\textbf{G}|} \sum_{g \in \textbf{G}}min_{p \in \textbf{P}} d(g, p)\ + \\
& \frac{1}{|\textbf{P}|} \sum_{p \in \textbf{P}} min_{g \in \textbf{G}} d(p, g)
\label{eq:11}
\end{aligned}
\end{equation} with $|\textbf{G}|$ and $|\textbf{P}|$ representing the number of voxels in the binary label maps of ground truth $\textbf{G}$ and prediction $\textbf{P}$, respectively. In contrast to the DSC, a smaller AVG HD indicates a better capture of the segmentation boundaries with a value of zero being the minimum (perfect match). Here, we use the AVG HD to evaluate the segmentation performance of the different network architectures against each other. 

\subsubsection{Intraclass correlation coefficient}
The intraclass correlation coefficient (ICC) is a widely used metric to assess both, the degree of correlation and agreement between measurements. Thus, it is an ideal metric to judge the reliability of a given method. The ICC ranges from 0 to 1, with values close to 1 representing high reliability. Here, we use the degree of absolute agreement among measurements also known as criterion-referenced reliability \cite{ICC} to compare the FastSurfer pipeline (deep learning segmentation + post-processing) to  FreeSurfer. To this end, we calculate the agreement between cortical thickness and subcortical volumes in consecutive scans using the OASIS1 test-retest set. Prior to ICC calculations, volume and thickness estimates of the subcortical and cortical structures are extracted from the segmented brains. After averaging across hemispheres, the ICC as well as the upper and lower bound with $\alpha=0.05$ level of significance are calculated for each region using the method described in \cite{ICC}. Additionally, cortical thickness maps are mapped to a common space (\textit{fsaverage}) and smoothed at 15 FWHM before calculating the ICC separately for each hemisphere. The resulting overlay maps are visualized on the semi-inflated \textit{fsaverage} surfaces.

\subsubsection{Group analysis}
A segmentation method can potentially reach high ICC while being insensitive to actual effects in the data. Therefore, it is important to validate the sensitivity of a given method with respect to its capability to detect known significant variations in brain morphology between diagnostic groups (group separability). We, therefore, fit identical linear models to FreeSurfer and to FastSurfer results each, explaining thickness or volume (dependent variable) by diagnosis controlling for age, sex, and (only for volume analysis) head size, using FreeSurfer's \textit{mri\_glmfit}. The p-values of the diagnosis effect are monotonically connected to the absolute value of the t-statistic (effect size divided by standard error) which in turn is a scaled version of Cohen's $d$, where the scaling factor depends on sample size. Given that here both methods operate on the same input images (and thus same sample sizes), a direct comparison of p-values across processing methods is meaningful, where a smaller p-value indicates a better group separability. 

These analyses are performed for volume of sub-cortical structures and average thickness of cortical regions of interest (ROIs), as well as for vertex-wise cortical thickness maps on both hemispheres. ROI measures are reported for structures that reach $p < 10^{-5}$ and are averaged across hemispheres for the sake of brevity. Cortical thickness is calculated as the minimal distance between the white matter and pial surface \cite{Fischl2000}. Prior to statistical analysis, thickness estimates of each subject were mapped to a common space (\textit{fsaverage}) and smoothed at 15 FWHM. Significance cut-off is set to p $<$ 0.05 without correction for multiple comparison as we are only interested in the relative differences between the two methods (FastSurfer and FreeSurfer). The p-value maps for each hemisphere are displayed on the semi-inflated \textit{fsaverage} surfaces. 

\section{Results}

\subsection{Accuracy}

In this section we evaluate segmentation performance with respect to FreeSurfer as a reference and with respect to manual labels. The comparison with FreeSurfer is relevant, given that the main goal of this work is to replicate FreeSurfer at faster processing speed without sacrificing quality.
Furthermore, the comparison with manual labels is important to validate accuracy and rank the methods with respect to a manual reference standard. 

\subsubsection{Comparison to FreeSurfer}
To evaluate segmentation accuracy, we first compute the DSC and AVG HD of each candidate method with FreeSurfer labels on five testsets (ADNI, OASIS1, HCP, MIRIAD, and THP). Here, scans from subjects used for network training and validation (i.e.\ 40 subjects each from OASIS and ADNI, 20 from MIRIAD) are excluded. No subjects from HCP and THP have been included in the training set at any point. Five subjects from OASIS1 were further excluded from the testset due to heavy white matter lesion load and resulting downstream topological surface defects in FreeSurfer and FastSurfer.

We benchmark our proposed network against traditional whole-brain segmentation F-CNNs namely SDNet~\cite{sdnet} and QuickNAT~\cite{quicknat}, as well as a patch-based 3D UNet. Additionally, we incrementally test the importance of our network modifications. First, we evaluate the effects of competition within the dense blocks and across the long-range skip connections (CDB, see Section~\ref{sec:competition}). For this we compare to QuickNAT which inherently uses only vanilla dense blocks within its architecture. Second, we increase the information input to the network by passing the stacked 7-channel image to the network (spatial information aggregation (SPI), see Section~\ref{sec:spatial_info}). Both architecture changes together comprise our final proposed FastSurferCNN. To permit a fair comparison, all benchmark networks were trained on the same data and - with exception of the 3D UNet - follow the same architectural design of 4 encoder and decoder blocks separated by a bottleneck block. Each block contains the same convolutional layer architecture as illustrated in Figure~\ref{fig:CNN_architecture}. The baseline architectures were further suitably adopted by modifying the final classification layer to predict 78 classes as the original implementations do not target cortical parcellations and hence comprise a much lower number of output labels (27 for QuickNAT \cite{quicknat} and 26 for SDNet \cite{sdnet}). Furthermore, all comparative 2D models were implemented with the above-mentioned view aggregation (see Section~\ref{sec:view_agg}). Care was taken to confirm that the adaptations are acceptable to the first author of the original papers.
The 3D UNet was optimized such that it allows segmentation at 1mm isotropic resolution with a patch size of 128x128x112, a batch size of 2, and 32 feature maps in the highest layer using a self-adapting framework \cite{nnUNet}. 

\begin{figure*}[!hbt]
    \centering
    \includegraphics[width=\textwidth, keepaspectratio]{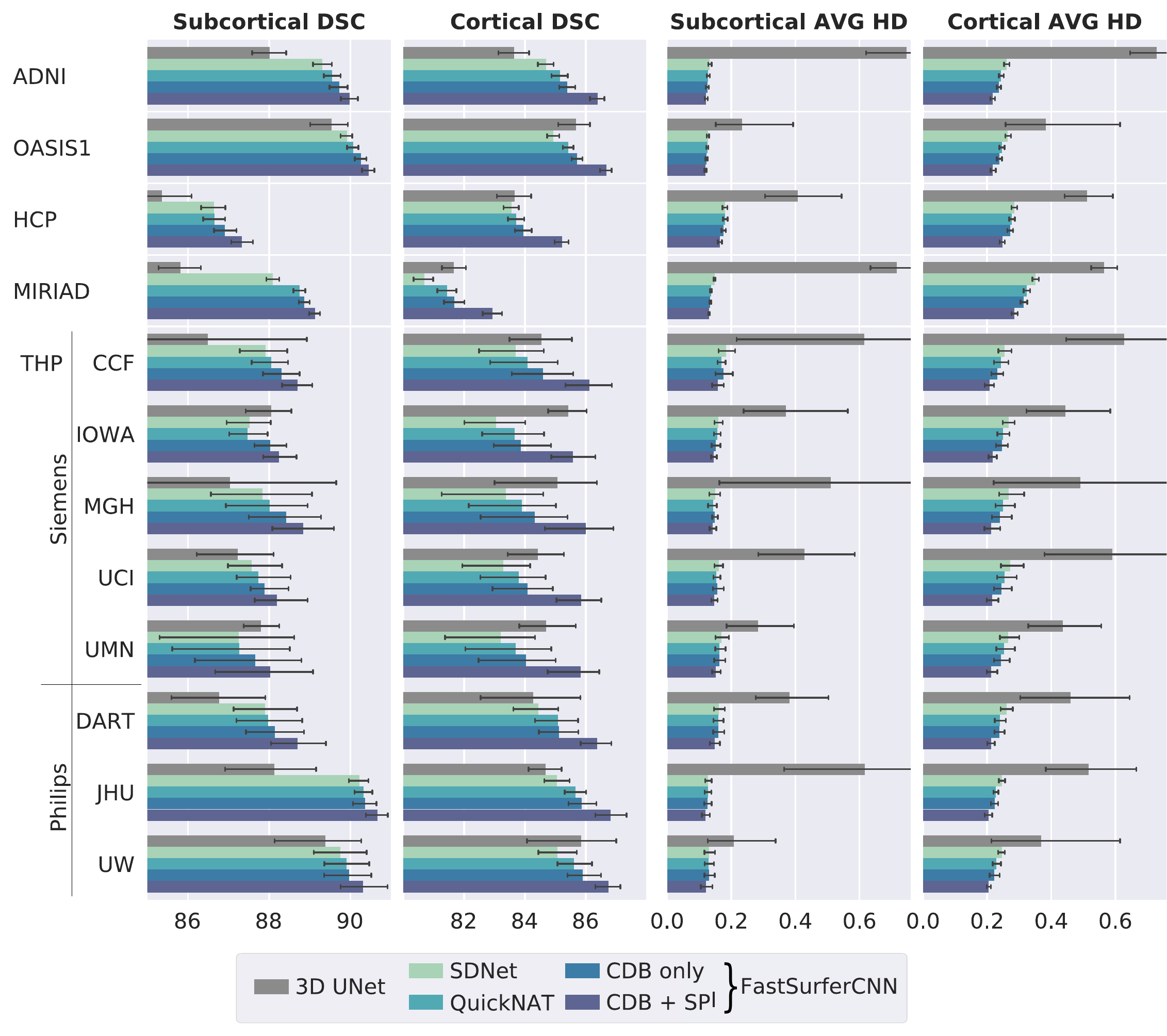}
    \caption{Dice similarity coefficient (DSC, left, larger = better) and average Hausdorff distance (AVG HD, right, smaller = better) comparison of baselines and the proposed FastSurferCNN on four different datasets (mean $\pm$ standard deviation). Network modifications (i) competitive dense blocks (CDB) and (ii) spatial information aggregation (SPI) are incrementally tested. The final FastSurferCNN (dark blue, CDB + SPI) outperforms all other models on both, subcortical and cortical structures.}
    \label{fig:hd_dice_plot_v2}
\end{figure*}

In Figure \ref{fig:hd_dice_plot_v2}, we report the DSC (left side) and AVG HD (right side) against 649 previously unseen scans from ADNI (180 subjects), OASIS1 (370 subjects), HCP (45 subjects), MIRIAD (49 subjects) and THP (5 subjects). We calculate the average DSC and AVG HD on 33 subcortical structures and 62 cortical regions (31 per hemisphere after remapping). For a complete list of structures the reader is referred to Appendix Table \ref{tab:labels}. The AVG HD should be as small as possible (optimal value: 0). An optimal segmentation in terms of DSC should be close to 100. 

In comparison to the 2.5D network approaches, the 3D UNet has the overall lowest DSC on the subcortical structures (average DSC of 87.27 - see Figure \ref{fig:hd_dice_plot_v2} left, gray bars), and is comparable only to the SDNet on the cortical structures (average DSC of 83.90, a 0.4 \% increase compared to SDNet, light green bars). However, it has by far the highest AVG HD (see Figure \ref{fig:hd_dice_plot_v2} right, gray bars) for both subcortical and cortical structures (0.5063 and 0.5357, respectively). The errorbars for the 3D UNet were cropped at 0.76 in the AVG HD plot.

Each successive network modification within the 2.5D network results in an increase of the DSC (see left part of Figure \ref{fig:hd_dice_plot_v2}) and a decrease of the AVG HD (see right part of Figure \ref{fig:hd_dice_plot_v2}) for all five datasets. Introduction of competition within the network (blue, CDB) already outperforms QuickNAT (green) and SDNet (light green) with an up to 0.3~\% improvement of the DSC and 5~\% of the AVG HD for both, subcortical and cortical structures. On average, competition increases the DSC to 88.74 and 84.55 for the subcortical and cortical structures, respectively. The AVG HD is decreased to 0.247 on the cortex and 0.1486 on the subcortex.
Note, that this improvement is achieved while simultaneously reducing the number of trainable parameters by one half (from approx.\ $3.6*10^6$ to $1.8*10^6$)!
The final FastSurferCNN (CDB+SPI, dark blue) further increases segmentation accuracy on average by 0.6~\% (DSC) and 5.7~\% (AVG HD) on the subcortical and 1.9~\% (DSC) or 12.7~\% (AVG HD) on the cortical structures compared to QuickNAT (final DSC of 89.08 and 85.88; AVG HD of 0.1400 and 0.2222). Therefore, increasing the local information content provided to the network via the spatial information aggregation (SPI) is particularly useful for recognizing cortical folding patterns. The same trend can be observed when analyzing the worst instead of the average DSC (data not shown). Statistical testing further confirmed a highly significant increase in DSC and decrease in AVG HD for both improvements (competition and information aggregation) compared to QuickNAT (Wilcoxon signed-rank test, $p < 10^{-20}$ after Bonferroni correction for multiple testing). 

FastSurferCNN also outperforms all other models on the challenging THP dataset. This data source contains scans from eight different sites and scanning conditions with strong variations in data quality (e.g.\ motion artifacts). FastSurferCNN, however, maintains high accuracy for all eight sites (average DSC 89.00, AVG HD: 0.1426 (QuickNAT: 88.37 and 0.1514, respectively) for subcortical and DSC: 86.16, AVG HD: 0.2136 (QuickNAT: 84.43 and 0.2465, respectively) for cortical structures). Interestingly, MRI data acquired on Philips Scanner (DART, JHU and UW) are segmented quite accurately even though the training set predominantly included Siemens scans. Additionally, the AVG HD and DSC are more stable across imaging sites for FastSurferCNN compared to QuickNAT on the cortical structures (40~\% less difference in DSC and 25~\% less difference in AVG HD between best (JHU) and worst (IOWA) site). Notably, the algorithm also generalizes well to defaced and downsampled input images (HCP dataset) even though such examples are absent from the training set (DSC of 85.21 for the cortical and 87.36 for the subcortical structures, AVG HD of 0.2474 and 0.1643, respectively).

\subsubsection{Comparison to Manual Reference}
Here we compare segmentation performance of the networks to a manual standard. We compute the DSC and AVG HD on the Mindboggle-101 dataset (78 subjects) with manually labeled cortical as well as manual subcortical segmentations on 20 of these subjects. 

\begin{figure*}[!hbt]
     \centering
    \includegraphics[width=\textwidth, keepaspectratio]{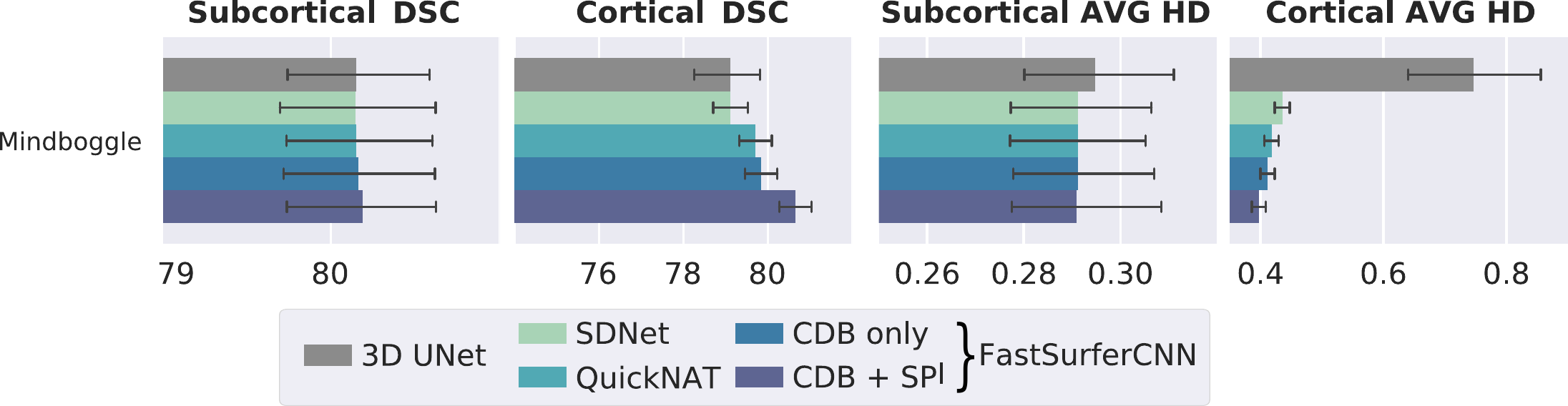}
     \caption{Dice similarity coefficient (DSC, left, larger = better) and average Hausdorff distance (AVG HD, right, smaller = better) comparison across networks with respect to a manual reference (Mindboggle-101). FastSurferCNN outperforms all other models on both subcortical and cortical structures.}
    \label{fig:mindboggle_eval_v2}
\end{figure*}

In Figure \ref{fig:mindboggle_eval_v2}, we report the DSC (left side) and AVG HD (right side) for the subcortical and cortical structures separately. The DSC on the subcortical structures are very similar for all five networks: between a DSC of 80.15 for 3D UNet, SDNet and QuickNAT and 80.19 for the final FastSurferCNN (CDB+SPI). On the cortical structures the difference is more pronounced, with the final FastSurferCNN (CDB+SPI) reaching a value of 80.65 (1~\% increase compared to CDB (blue) and QuickNAT (green), 2~\% compared to SDNet (light green) and 3D UNet (gray)). The same trend is observed for the AVG HD (right plot). Here, introduction of competition (CDB) decreases the distance between segmentation and the manual labels by 0.025~\% on the subcortical and 1.56~\% on the cortical structures (CDB (blue) versus QuickNAT (green)). The final FastSurferCNN (CDB+SPI, dark blue) reaches a value of 0.2909 for the subcortical (0.085~\% improvement compared to SDNet (light green) and QuickNAT (green)) and 0.3973 on the cortical structures (~10~\% and 5~\% compared to SDNet and QuickNAT, respectively). The 3D UNet is again the lowest performing out of all the five networks with an AVG HD of 0.3443 for the subcortical and 0.7472 for the cortical structures, indicating large inaccuracies in especially the cortical regions. 

The improved segmentation performance of FastSurferCNN on the manual cortical labels with respect to DSC and AVG HD was further confirmed by statistical testing (Wilcoxon signed-rank test, $p < 10^{-10}$ after Bonferroni correction for multiple testing).

\begin{figure}[!hbt]
     \centering
    \includegraphics[width=\columnwidth,keepaspectratio]{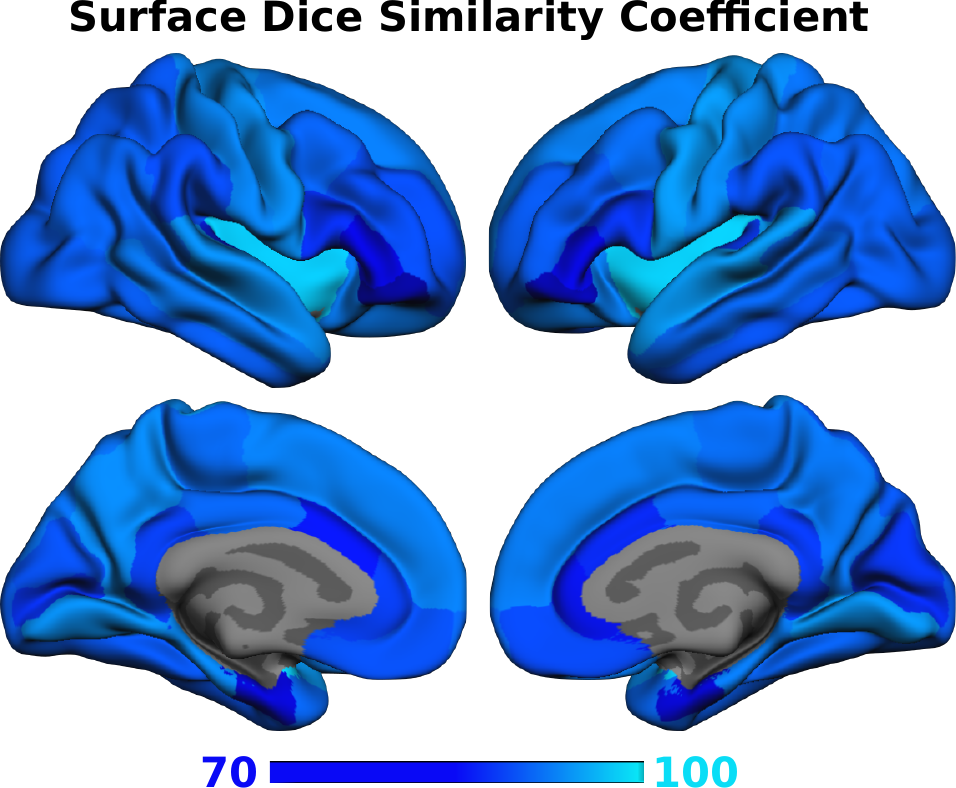}
     \caption{Surface Dice similarity coefficient per region of the proposed FastSurferCNN for the 31 cortical parcels on manual surface labels from Mindboggle-101. FastSurferCNN achieves accurate results across structures with an average DSC of above 86.}
    \label{fig:surface_dsc}
\end{figure}

Inherently, cortical regions are usually defined on the cortical surface based on folding patterns and curvature, e.g.\ to determine boundaries inside the sulcii. Thus, assessment of cortical segmentation quality may not be optimal in a volume based comparison. Therefore, we also calculate the surface-based DSC for Mindboggle-101 (manual surface labels). To this end, the volumetric segmentation of FastSurferCNN are projected onto the Mindboggle-101 surfaces where the area-related DSC of the mapped regions are then directly calculated. Overall, a high average surface DSC of 86.35 on the right and 86.97 on the left hemisphere is reached. Further, no structure has a DSC below 72.3 (see Figure \ref{fig:surface_dsc}). The good performance of the surface analysis corroborates the volume based DSC comparisons and underlines the high segmentation quality achieved with FastSurferCNN. 

\subsection{Generalizability}

High generalizability will ensure that the proposed method can be applied across different sites, vendors, field strengths, and for large multi-center studies. Figure \ref{fig:hd_dice_plot_v2} indicates that networks generalize well across these parameters and respective image qualities, as the DSC remains quite stable. For example, the HCP dataset consists of 0.7~mm isotropic images, downsampled to 1~mm and de-faced, which were never encountered during training. MIRIAD is a IR-FSPGR sequence on a GE scanner. Furthermore, in the THP dataset DSC scores vary only around 1 or 2~\% across the 8 sites spanning Siemens and Philips scanners. The five subjects of THP, however, might not be representative, which is why in this section we quantify generalizability by computing the agreement of FastSurferCNN with FreeSurfer across different scanner types (Siemens, Philips and GE) as well as disease states (CN, MCI and AD patients) in a larger dataset. For this purpose we employ an independent testset consisting of 180 scans from ADNI balanced with regard to vendor, disease group, gender and age. 
\begin{figure}[!hbt]
     \centering
     \includegraphics[width=\columnwidth,keepaspectratio]{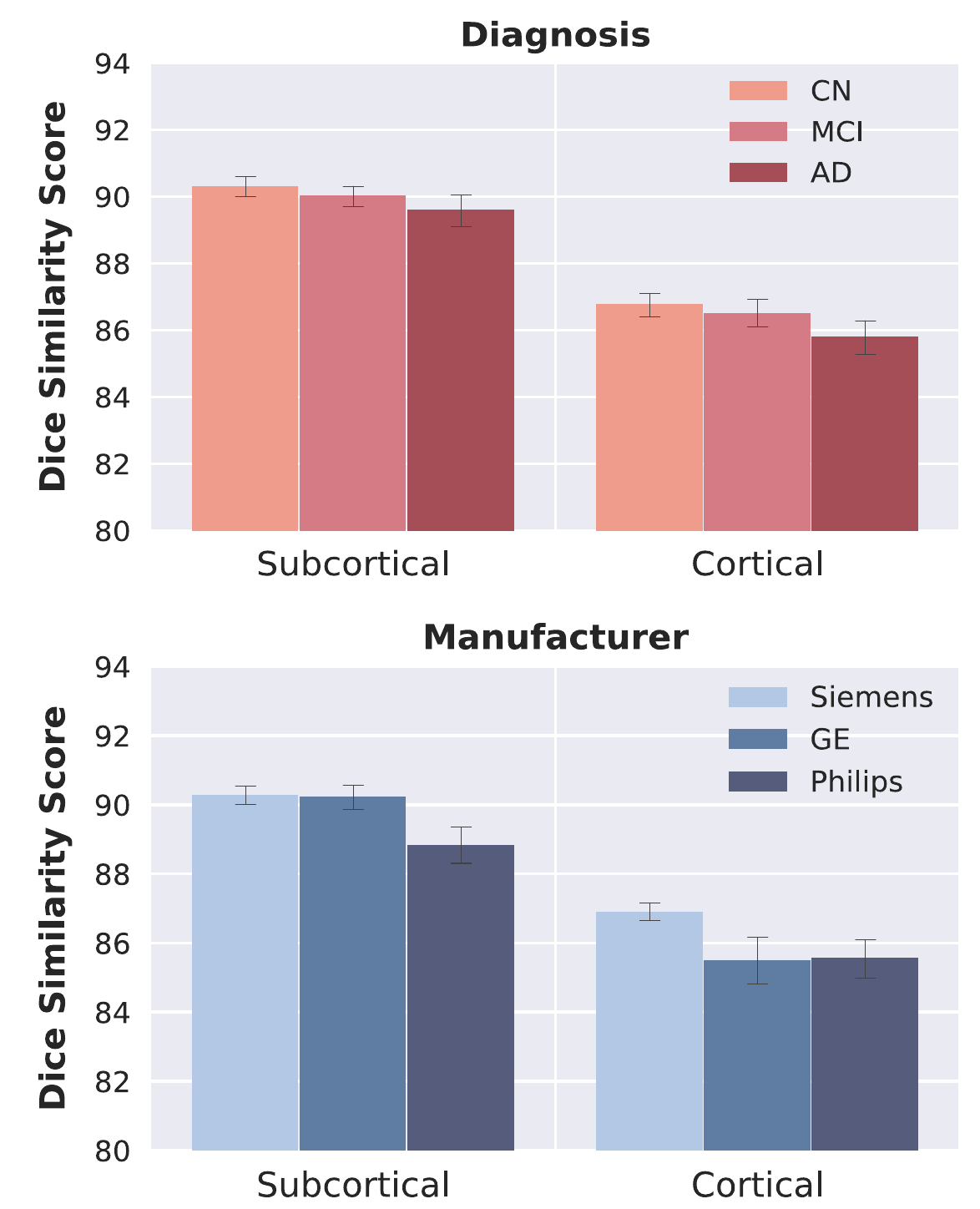}
     \caption{Comparison of the DSC (mean $\pm$ standard deviation) across neurodegenerative states (cognitive normal (CN), mild cognitive impaired (MCI), demented (AD); top) and vendor (GE, Philips, Siemens; bottom). FastSurfer achieves high accuracy and low variability across all of them.}
   \label{fig:dsc_diagnosis_scanner}
\end{figure} 

In the upper part of Figure~\ref{fig:dsc_diagnosis_scanner} the DSC across different disease states is shown. Scans from subjects with later-stage dementia are more difficult to segment, potentially due to increased motion and lower gray-white contrast. Specifically, the increase in ventricle volume, subsequent shrinkage of GM and/or increased white matter lesion load can have a profound effect and are frequently difficult to segment with traditional neuroimaging pipelines such as FreeSurfer. A small decrease in segmentation DSC (i.e.\ increased deviation from FreeSurfer) can be observed with disease progression (CN to AD) in our proposed method (cortical structures: DSC decreases from 86.8 to 85.8, subcortical structures: DSC decreases from 90.3 to 89.6). However, with FastSurferCNN DSC scores do not decrease more than 1.11~\% between diseased (MCI, AD) and control state (CN) for cortical and 0.77~\% for subcortical structures. A similar trend is observed with regard to generalizability across vendors (lower part of Figure~\ref{fig:dsc_diagnosis_scanner}). Even though FastSurferCNN was predominantly trained on MRI scans acquired on Siemens scanners, the segmentations of Philips and GE scans are comparable to FreeSurfer with only a minor decrease in DSC of 1.6~\% for the cortical structures (see Figure~\ref{fig:dsc_diagnosis_scanner}). Here, GE and Philips scans can be segmented similarly well (DSC of 85.50 and 85.57, respectively). Segmentation of subcortical regions is equally consistent on Siemens and GE scans (DSC of 90.30 and 90.23, respectively). Philips scans show more variation with a difference similar to the one observed for the cortical structures (1.6~\% difference compared to Siemens; DSC of 88.85). 
These findings in combination with the high DSC scores achieved on MIRIAD (1.5T GE Scanner) and HCP (downsampled and defaced 0.7~mm 3T Siemens), see Figure~\ref{fig:hd_dice_plot_v2}, show that FastSurferCNN is capable of generalizing stably across disease stages, field strength, vendors and pre-processing. 

\begin{figure}[!hbt]
     \centering
     \includegraphics[width=\columnwidth,keepaspectratio]{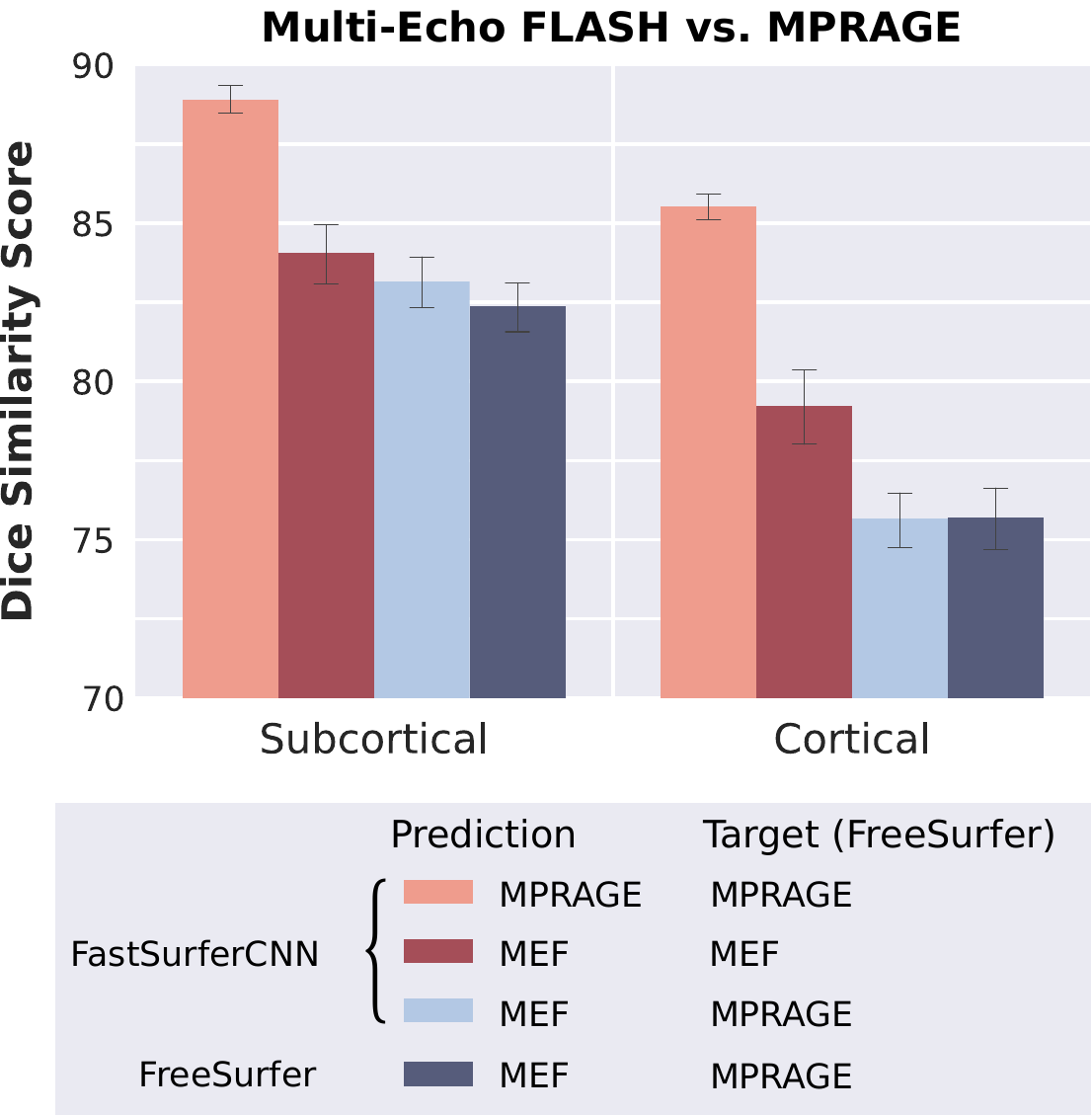}
     \caption{Comparison of segmentation accuracy between different image acquisition sequences (MPRAGE versus Multi-echo FLASH (MEF)). FreeSurfer's MPRAGE segmentations are considered ground truth. FastSurferCNN generalizes well to MEF acquired images and achieves closer results to FreeSurfers MPRAGE than FreeSurfer itself on the subcortical structures.}
     \label{fig:flash_plot}
\end{figure}

While very few GE scans were used in training (only 3 cases of ADNI), still the network has at least seen some images spanning the above parameters which might explain its good generalizability. All datasets so far acquire some kind of MPRAGE sequence, except for MIRIAD with IR-FSPGR (note, MIRIAD was not used in training, only for validation). We now test generalizability across sequences to an unseen MEF sequence  (MMND dataset, see Section \ref{sec:datasets}). This datasets provides 16 subjects each with MEF and MPRAGE scans. 
We first confirm, see Figure~\ref{fig:flash_plot} (orange column), that FastSurferCNN obtains a high DSC in comparison to FreeSurfer on the MPRAGE images, again corroborating good generalizability to Philips scanners. A 6.7~\% point drop in DSC can be observed when comparing both segmentation methods on the MEF images (red column). Reduced agreement on MEF images can, of course, be explained by a deviation of either (or both) methods from ground truth. In the absence of ground truth, we set FreeSurfer's MPRAGE segmentation as the standard. To test how well FreeSurfer generalizes to MEF, we compare FreeSurfer's own segmentations across  MEF and MPRAGE after robust rigid registration \cite{reuter2010robustreg} and observe a significantly reduced DSC (82.37 on subcortical and 75.70 on cortical structures) (Figure~\ref{fig:flash_plot} dark blue). In comparison FastSurferCNN's MEF segmentation (using the same registrations) slightly outperforms FreeSurfer's generalizability as it is actually closer to ground truth (FreeSurfer MPRAGE) than FreeSurfer itself for subcortical structures (DSC of 83.17) and similar for the cortex (DSC of 75.67) (Figure~\ref{fig:flash_plot} bright blue). These results highlight an excellent generalizability of FastSurfer to the unseen T1-weighted MEF sequence. 

The above DSC comparisons assume accurate labels provided by FreeSurfer or a human rater respectively. This is of course not granted, as both automated and manual segmentation quality can degrade across scanners, sequences, or advanced neurodegeneration. Furthermore, highly folded 3D structures such as cortical gray matter are difficult to segment manually on a voxel grid when viewing 2D slices. 
Whenever ``ground truth'' cannot be trusted, it is difficult to quantify performance with direct comparisons, as a small DSC can also indicate noisy or erroneous ground truth. Therefore, we also perform validations in the next sections that are independent of ground truth labels but rather rely on the assumptions, that (i) anatomy remains stable in small time frames (test-retest reliability) and that (ii) established disease effects should be detectable with high statistical significance (sensitivity). For these comparisons we run the full FastSurfer pipeline, extending the FastSurferCNN with subsequent surface processing.

\subsection{Reliability}

\textbf{Test-retest reliability} is assessed as the agreement between the evaluations of two scans in a short time frame. We calculate the intraclass correlation coefficient \cite{ICC} on the OASIS1 test-retest dataset with 20 participants. Note, that the acquisition source of variation (motion, noise etc.) will be identical for different image processing methods. Higher agreement can therefore be taken as an index of method stability and consistency of results.

\begin{figure}[!hbt]
    \centering
    \includegraphics[width=\columnwidth,keepaspectratio]{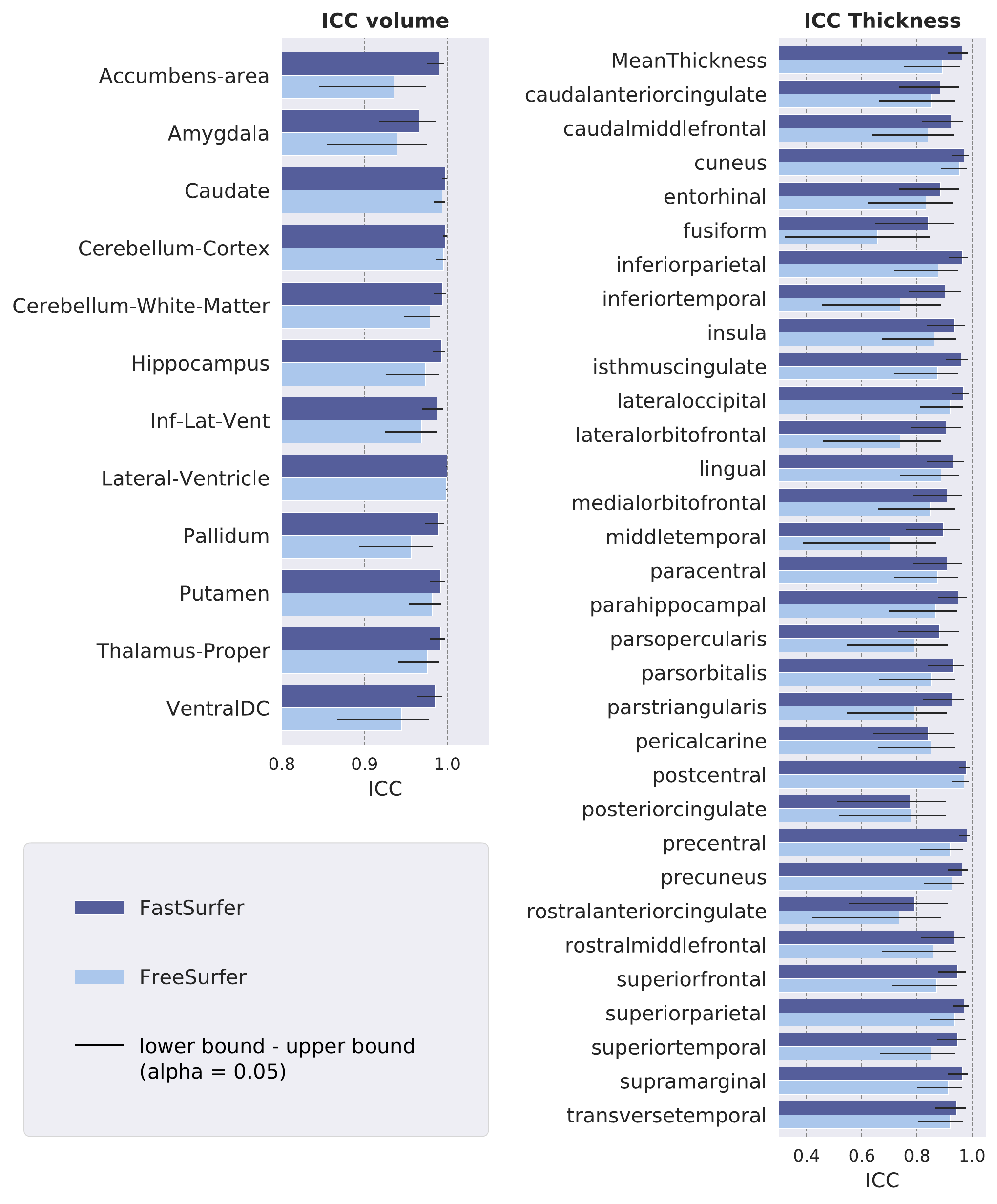}
    \caption{Intraclass correlation coefficient on the Test-Retest OASIS1 dataset for FastSurfer (dark blue) and FreeSurfer (light blue). Error bars indicate upper and lower bound of the calculated ICC (significance level $\alpha$=0.05).}
    \label{fig:oasis_icc_plot}
\end{figure}

\begin{figure}[!hbt]
    \centering
    \includegraphics[width=\columnwidth,keepaspectratio]{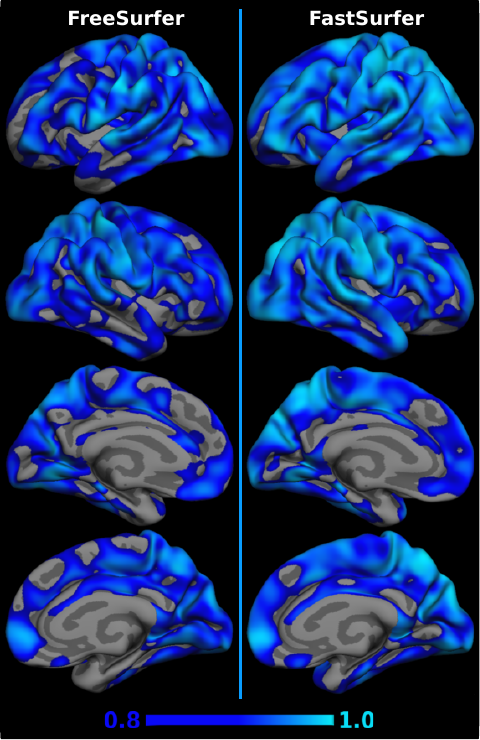}
    \caption{Visualization of the intraclass correlation coefficient on the Test-Retest OASIS1 dataset for FreeSurfer (left) and FastSurfer (right). ICC ranges from 0.8\ (dark blue) to 1.0\ (light blue) are shown.}
    \label{fig:oasis_icc}
\end{figure}

Figure \ref{fig:oasis_icc_plot} shows the ICC value for each structure separately including the upper and lower bound at significance level $\alpha$=0.05 (black error bar) for FastSurfer (dark blue bars) and FreeSurfer (light blue bars). A higher ICC indicates a better reliability with the maximum being 1. The ICC values for the volume of subcortical structures are higher for FastSurfer on all 13 structures (0.99 on average versus 0.97 with FreeSurfer). On six structures (including Hippocampus, Putamen, Caudate and Thalamus) the agreement between scans is above 0.99. Furthermore, all confidence intervals (lower to upper bound) are smaller for FastSurfer indicating better segmentation consistency across the 20 participants. The ICC values for the thickness of cortical regions show a similar pattern, with FastSurfer achieving higher ICC on both hemispheres in 30 out of 31 regions. On average, an ICC of 0.92 is achieved with FastSurfer (0.87 with FreeSurfer) and 24 structures show a correlation of above 0.9. No cortical structure has an ICC below 0.78 (0.73 for FreeSurfer). Correspondingly, visualization of the ICC directly on the surface (see Figure~\ref{fig:oasis_icc}) demonstrates that FastSurfer segmentations yield larger regions on the cortex with high ICC values (light blue) compared to FreeSurfer. Here, it is further apparent that the majority of the cortex reaches ICC values of more than 0.8\ (blue areas). 

\subsection{Sensitivity to Group Differences in Dementia}

\textbf{Sensitivity} to group differences in dementia is an essential component of our evaluation. While the segmentation accuracy of our approach (as quantified by the DSC to a manual standard) has already been established, we now determine to what extent our results are relevant and useful in an applied research setting. To this end, we analyze whether our proposed method is capable of reproducing well-known group differences in dementia: 
The sensitivity of both FreeSurfer and FastSurfer is determined by evaluating their ability to separate diagnostic groups in OASIS1 (AD versus CN), as indicated by the p-value.

\begin{figure}[!hbt]
    \centering
    \includegraphics[width=\columnwidth,keepaspectratio]{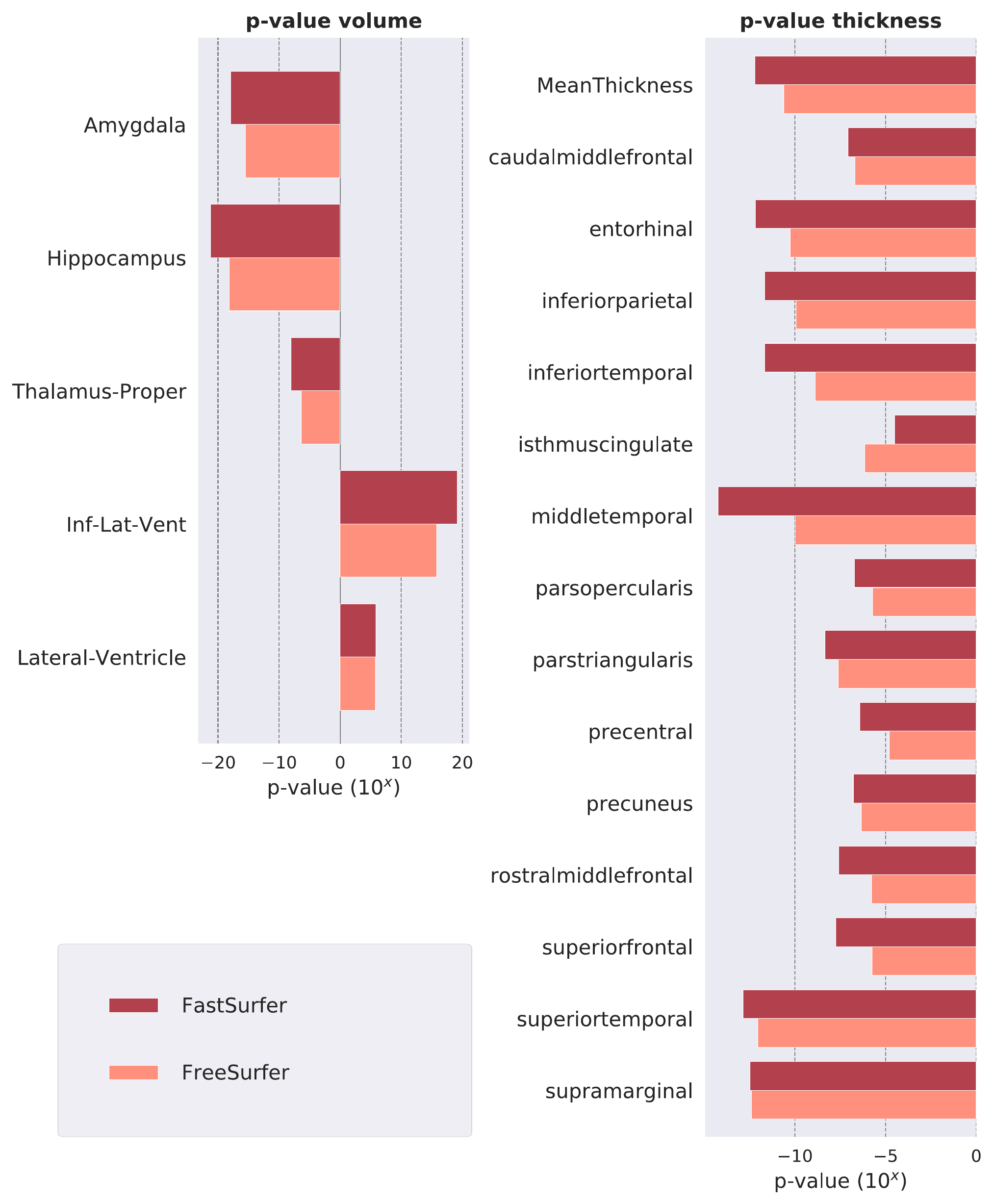}
    \caption{Significance (p-value) of cortical thickness changes in disease groups on the OASIS1 dataset for FastSurfer (red) and FreeSurfer (light red). Structures with p-values below $10^{-5}$ are shown. Effect directions are indicated by the sign (atrophy: negative values, enlargement: positive values). Volume of ventricles increases while all other structures show atrophy. FastSurfer and FreeSurfer detect significant changes in areas related to disease progression.}
    \label{fig:oasis_group_plot}
\end{figure}

In Figure \ref{fig:oasis_group_plot} structures with p-values below $10^{-5}$ are shown (based on both FreeSurfer and FastSurfer). The signed p-values presented in the figure indicate the direction of the effect (below zero represent atrophy and above zero volume increase). Consequently, one can directly observe that the ventricle volume (lateral ventricles, inferior lateral ventricles) increases while all other structures atrophy for both, FreeSurfer and FastSurfer. In the subcortical domain, a volume reduction is specifically detected for the hippocampus, the amygdala and the thalamus, which is congruent with other research results on AD \cite{deJong2008, Henneman2009, Schuff2009, Poulin2011, Aggleton2016, Pini2016}. FastSurfer reaches lower p-values for all three structures indicating a higher sensitivity to differences between the groups. 
FastSurfer and FreeSurfer are further capable of detecting significant differences in areas related to disease progression in the cortex (e.g.\ bilateral frontal, temporal and parietal lobe; \cite{Braak1995, Baron2001, Wenk2003, Lerch2005, Poulin2011}). Specifically, parts of the temporal (superiortemporal, middletemporal, inferiortemporal, entorhinal) and parietal lobes (inferiorparietal, supramarginal) are significantly thinner (p $<10^{-11}$ detected with FastSurfer for all areas). The overall thickness also correlates with disease progression (MeanThickness p $<10^{-12}$ for FastSurfer). These results remain stable when analyzing lateralized ROI measures instead of the mean (not shown). 

\begin{figure}[!hbt]
    \centering
    \includegraphics[width=\columnwidth,keepaspectratio]{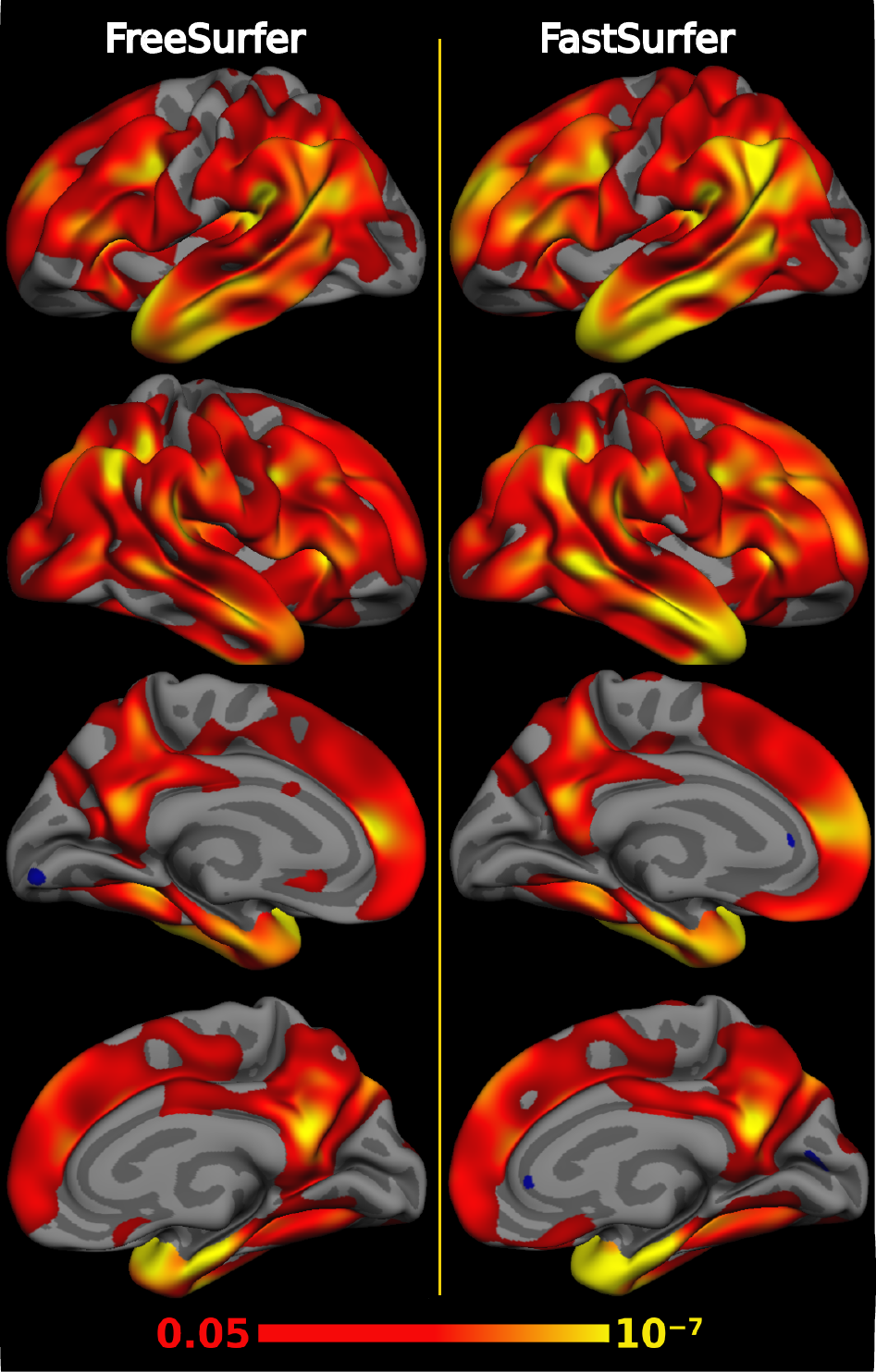}
    \caption{Group Analysis of cortical thickness variations in Alzheimer’s disease compared to controls based on the OASIS1 dataset for FreeSurfer (left) and FastSurfer (right). The color-coded uncorrected p-value map ranges from 0.05\ (red) to $10^{-7}$ (yellow). Differences in cortical thickness are more pronounced in the FastSurfer analysis stream.}
    \label{fig:oasis_group}
\end{figure}

Figure \ref{fig:oasis_group} depicts the detected differences of cortical thickness in patients with AD compared to CN subjects with the original FreeSurfer stream (left) and our proposed FastSurfer pipeline (right) directly on the surface (vertex-wise analysis). The visualization complements the results of Figure~\ref{fig:oasis_group_plot}, clearly indicating the ability of FreeSurfer and FastSurfer to detect thinning effects across hemispheres. Again, the differences between groups are more pronounced with the FastSurfer pipeline (smaller uncorrected p-value, as indicated by larger yellow regions on both hemisphere). The proposed pipeline is thus able to separate groups in this dataset very clearly. 

\subsection{Pipeline Innovations}

As described in Section \ref{sec:pipeline}, FastSurfer is a full surface reconstruction pipeline based on FreeSurfer. Additionally to skipping/replacing some steps the two main modifications are (i) reconstructing surfaces with the marching cube algorithm and (ii) implementing a novel, fast spectral mapping to the sphere. Here we compare these two changes with a pipeline that uses the original FreeSurfer modules (\textit{mri\_tessellate} and \textit{mris\_sphere}) in these two steps and is otherwise identical to FastSurfer. We quantify a) the number of topological surface defects, b) the overall processing time for the surfaces, and c) the average quality of the produced surface triangle meshes. 
Surface meshes should be topologically equivalent to the sphere, which is why topological defects introduced during surface construction, i.e.\ handles and tunnels, as well as self-intersections have to be corrected in time-intensive downstream processing \cite{Fischl1999a, Segonne2007}. This is why a smaller number of topological defects on the original meshes is favourable.
The mesh quality is computed by averaging triangle qualities $Q = 4 \sqrt{3} \ A / (e_1^2 + e_2^2 + e_3^2 )$, where $A$ is the triangle area and $e_i$ the edges \cite{Bank90}. $Q$ is 1 for the equilateral triangle and close to zero for degenerated triangles.

To evaluate the two new modules, the following statistics are based on the OASIS1 dataset. The average number of defects in the pipeline with FreeSurfer modules (27.2 defects per hemisphere) decreased by 12~\% (24.0 defects) when constructing the surfaces with marching cube, and 15.3~\% (23.1 defects) when the proposed pipeline (FastSurfer: marching cube + spectral spherical projection) was used. Also the average surface processing time decreased significantly by 15 minutes per hemisphere with FastSurfer. Finally the triangle mesh quality index of FastSurfer (Q=0.902) was significantly higher than for the pipeline with the traditional modules (Q=0.888), due to the marching cube algorithm. Additionally, construction of surfaces with marching cube instead of \textit{mri\_tessellate} slightly reduces the number of vertices on average by 120 per hemisphere (133154 versus 1333034 for tessellate and marching cube, respectively). All three tests (defects, time, quality) of the final FastSurfer pipeline with respect to the one with original modules are highly significant: the Wilcoxon signed-rank test reports a p-value below machine precision (p $< 10^{-16}$).

\begin{table}[]
\caption{Average number of topological defects, mesh quality and processing time per hemisphere when using original FreeSurfer modules, marching cube, and spectral spherical projection in the FastSurfer pipeline.}
\centering
\begin{tabular}{|l|c|c|c|l}
\cline{1-4}
                      & \multicolumn{1}{l|}{\textbf{Defects}} & \multicolumn{1}{l|}{\textbf{Quality}} & \multicolumn{1}{l|}{\textbf{Time/hemi}} &  \\ \cline{1-4}
\textbf{orig}         & 27.2                                  & 0.888                                   & 41.6                                    &  \\ \cline{1-4}
\textbf{+ mc}         & 24.0                                  & 0.902                                   & 25.7                                    &  \\ \cline{1-4}
\textbf{+ qspec} & 23.1                                  & 0.902                                   & 25.4                                    &  \\ \cline{1-4}
\end{tabular}
\label{tab:pipeline_comp}
\end{table}

Overall, in addition to these methodological innovations, our pipeline saves time by replacing many FreeSurfer steps, such as skull stripping, spherical segmentation etc.,\ since we can directly build on the high-quality image segmentations provided by the deep neural network. Here, we compare runtimes of three approaches: (i) complete regular FreeSurfer processing, (ii) FastSurfer pipeline without spherical registration, (iii) and complete FastSurfer with spherical registration. Note, while spherical registration is not needed to obtain surface segmentations and ROI thickness measures in FastSurfer, it is required to construct cross-subject correspondence, e.g.\, when performing vertex-wise surface thickness analyses. All pipelines are evaluated on identical subjects (10 representative subjects from OASIS1, balanced with regard to gender (5 male, 5 female), age range (21 to 86 years), and diagnosis (4 AD, 6 CN)) and identical hardware (CPU: Intel Xeon Gold 6154 @ 3 Ghz) using both, sequential processing and parallel processing (4 threads and simultaneous processing of the two hemispheres).

In our test a complete FreeSurfer run takes approximately 7h (4h parallel) on the CPU, which can vary depending on image quality, disease severity etc. Obtaining only the  {\it aseg.mgz} (36 labels including subcortical structures) takes approximately 2-3 h for FreeSurfer.
However, improved GM and WM segmentations as well as cortical segmentation (DKT Atlas) is only available after surface reconstruction and remapping to the volume as one of the final steps. Thus, obtaining the final segmentation takes the full processing time (7h, 4h parallel). The proposed FastSurfer pipeline achieves the volumetric segmentation (both subcortical and cortical regions) in only 1 minute (on the GPU, 14 min on the CPU), surface processing including cortical ROI thickness measures in 1.7h (0.9h parallel), or complete processing including the spherical registration for potential subsequent group analysis of surface maps in 3.7h (1.6h parallel) on the CPU.

\begin{table}[]
\caption{Runtime comparisons of the different pipelines (recon-all, recon-surf and recon-surf including registration (reg) of the orig surface to the spherical atlas for cross-subject correspondence). Average processing times and standard deviation in minutes. Parallel = parallelization of hemispheres, 4 threads}
\centering
\begin{tabular}{|l|c|c|}
\hline
                                 & \textbf{sequential} & \textbf{parallel} \\ \hline
\textbf{FreeSurfer}               & 424  min ($\pm$ 37)      & 244 min ($\pm$ 20)    \\ \hline
\textbf{FastSurfer}              & 104 min ($\pm$ 20)      & 54 min ($\pm$ 27)     \\ \hline
\textbf{\begin{tabular}[c]{@{}l@{}}FastSurfer + \\ sphere.reg\end{tabular}} & 223 min ($\pm$ 53)      & 97 min ($\pm$ 31)     \\ \hline
\end{tabular}
\label{tab:runtimes}
\end{table}

\section{Discussion}

In this work we introduce an advanced semantic segmentation neural network architecture and a fast pipeline for the processing of neuroanatomical surfaces that outperforms FreeSurfer with respect to runtime, reliability and sensitivity. 
We contribute a deep learning architecture (FastSurferCNN) by incorporating competition into the network (within dense blocks and long-range skip connections) and increasing the initial information content provided to the network via spatial information aggregation. Competition significantly reduces the number of network weights resulting in a slimmer architecture with lower memory requirements. We demonstrate its superior performance for the fast and detailed (close to 100 structures) segmentation of whole brain MRI compared to existing deep learning approaches. FastSurferCNN outperforms 3D UNet, SDNet and QuickNAT in terms of accuracy by a significant margin both with respect to FreeSurfer and a manual standard. Across five different datasets our network achieves the highest DSC compared to FreeSurfer as a reference on the subcortical and cortical structures (89.08\ and 85.87\ on average), as well as the lowest AVG HD (0.1400 and 0.2222 on average). In addition, FastSurferCNN is the best performing network on the manually labeled Mindboggle-101 dataset with a DSC of 80.19 and 80.65 and an AVG HD of 0.2909 and 0.3973 on the subcortical and cortical structures, respectively.

Of note, the optimized 3D UNet was not able to outperform the other view-aggregating (2.5D) architectures. Specifically in terms of the AVG HD, the 3D UNet demonstrated underwhelming performance. In general, it can be expected that full 3D approaches eventually outperform 2.5D networks for whole brain segmentation due to their potential to better capture the inherent 3D geometry of the cortex. The better performance of the 2.5D approaches here might in part be explainable by their advanced architectures which are difficult to directly translate to a full 3D approach due to the increase in parameters and hence memory requirements. Furthermore, adding spatial information (SPI) as well as view-aggregating the three 2D networks provides spatial context at a much lower memory prize and training effort. 
In depth comparison of 2.5D versus 3D networks as well as improvements of 3D architectures is however out of scope of this work and an important topic for future exploration.

Further, FastSurferCNN's segmentation results for a 3D 1mm isotropic MRI brain scan are achieved at a processing time below one minute. Fast MRI segmentation opens up multiple avenues of potential applications, ranging from direct feedback or field-of-view localization during image acquisition, fast clinical decision support by quantitative personalized measurements, and scalability to very large cohort data sets. Many such applications require no surface models and can terminate after the 1 minute image segmentation step, allowing rapid processing of the incoming data.

One frequently quoted limitation of learning-based approaches is the uncertain generalizability beyond image types encountered during training. This limitation is valid and it remains unclear how far networks generalize, e.g.\, to different sequences, disease or age groups. Various domain-adaptation approaches have been introduced to accommodate fine-tuning a network to a new type of data, and should be considered when network performance degrades. In this work, we put an emphasis on evaluating generalizability of our method. We first demonstrate good generalizability to different sites, vendors, field strength, scanner types, and across disease groups. Analysis on HCP further highlights generalizability to de-faced and down-sampled high-resolutional images, which were never encountered during training. Furthermore, we were able to demonstrate good generalizability to an unseen multi-echo FLASH sequence, even outperforming FreeSurfer. These results are very promising, yet, as with any automated software, we recommend users to visually inspect images to ensure good quality for their acquisition setting as stable generalizability to any T1-weighted sequence can certainly not be guaranteed. 

Extending the image segmentation network, our full FastSurfer pipeline permits the fast analysis of cortical thickness (vertex-wise and region-wise) following the DKT atlas. This is achieved by both optimizing and replacing multiple steps of the FreeSurfer pipeline, e.g.\ by mapping segmentation results from the image to the surfaces. Processing of a single MRI volume with parallelization can thus be achieved in below 1h including thickness ROI analysis, and 1.6h including surface registration for cross-subject correspondence - a fraction of the time a whole FreeSurfer run needs to complete (4h with parallelization). Some of this speed-up can also be attributed to the reduced number of detected topological defects. Marching cube seems to be reducing the number of defects on the initial surfaces already, but also the new spectral spherical mapping helps further reduce detected defects, potentially due to the smooth embedding of the Eigenfunctions and resulting reduction of self-folds. Future work will focus on increasing processing speed further, e.g.\, by including deep-learning based registration procedures. Also note, that ongoing activities to parallelize and speed-up traditional FreeSurfer code will directly impact multiple components of the FastSurfer pipeline (such as the topology fixer, surface reconstruction, cross subject registration etc.).

Our extensive validation of FastSurfer further includes test-retest reliability and sensitivity studies. FastSurfer exhibits improved test-retest reliability relative to FreeSurfer. This is reflected in higher ICC values across both hemispheres for FastSurfer (on average 0.92 for all cortical and 0.99 for all subcortical structures). Given that increased reliability can be bought by extensive smoothing, potentially at the cost of sensitivity, we evaluate FastSurfer's capability to separate groups in dementia.
Here, we can replicate group differences between CN and AD patients with high sensitivity. Specifically, AD-related significant volume reductions in amygdala and hippocampus, increased ventricle volume, as well as cortical thinning in the temporal and parietal lobes were detected with both FastSurfer and FreeSurfer. 

While these group differences may not purely reflect neurodegeneration, but can include indirect factors such as head motion \cite{ReuterAndTisdall2015,TisdallAndReuter2016} or hydration levels \cite{Biller2015} the good agreement of FastSurfer with FreeSurfer and with established findings \cite{Braak1995, Baron2001, Wenk2003, Lerch2005, deJong2008, Henneman2009, Schuff2009, Poulin2011, Aggleton2016, Pini2016} indicate the validity of our method. FastSurfer's smaller p-values in the majority of these discrimination tasks could potentially be explained by its implicit noise reduction: While consistent boundaries in FreeSurfer segmentations will be learned, random segmentation noise such as local inaccuracies or protrusions are averaged out and might allow the network to achieve superior results, as also corroborated by the improved test-retest reliability results. 

The inherent training paradigm of FastSurferCNN can be considered another contributing factor. During training, the network has been exposed to various pathological scans with high anatomical and acquisition variability in contrast to the limited number of cases (40) within the FreeSurfer atlas. The larger corpus likely improves the resulting segmentations and derived volume and thickness estimates. In fact, it is remarkable that the 140 training cases (plus augmentation) are sufficient to provide these excellent results. This is put into perspective by considering that in fact 20k 2D images (plus augmentation, some of these highly correlated - of course) are used for training each view. Still it can be expected that training with more cases will improve accuracy and generalizability further, leaving space for future exploration.

Finally, one of the major advantages of supervised learning over traditional pipelines is that consistent errors can be removed by manually fixing existing or adding new training cases. This is in stark contrast to model based pipelines, where updates or fixes to the algorithm can only be introduced by a handful of experts and often have unintended consequences. Future work can thus explore training on very large and heterogeneous datasets, as well as the inclusion of manual labels or manually corrected automated labels to improve segmentation quality even further. 

It should, however, be noted that availability of manually labelled or corrected training data is often inherently limited due to time intensive production, restricting the applicability of deep-learning approaches to domains were enough training data exists. This is especially true for full 3D segmentation networks were one whole brain MRI is equivalent to a single training case. 2D networks operating on slices or 3D patch approaches are more forgiving due to the much larger number of training inputs. In this work, training on FreeSurfer generated outputs allows us to generate a wide range of training cases with the disadvantage that consistent FreeSurfer segmentation errors may be learned by the networks. Our successful comparison to a manual reference standard (Mindboggle-101) is, therefore, helpful to ascertain high segmentation quality. Yet, it does not replace quality control when applying FastSurfer to new images, as is the case with any automated processing method.

Overall we introduce a fast, stable, reliable and sensitive pipeline for automated neuroimage analysis that scales well to large datasets and enables various new applications where segmentation speed is essential, for example: to localize structures during image acquisition, to provide quantitative measures in clinical workflows, or to process large cohort studies efficiently.

\section{Acknowledgment}

Support for this research was provided in part by the Federal Ministry of Education and Research of Germany (FKZ: 031L0206 CompLS2 DeepNI), by NIH R01NS0\-83534, R01\-LM012719, and by an NVIDIA Hardware Award as well as the BRAIN Initiative Cell Census Network grant U01MH117023, the National Institute for Biomedical Imaging and Bioengineering (P41EB015896, 1R01EB023281, R01EB006758, R21EB018907, R01EB019956), the National Institute on Aging (1R01AG064027, 1R56AG064027, 5R01AG008122, R01AG016495), the National Institute of Mental Health  the National Institute of Diabetes and Digestive and Kidney Diseases (1-R21-DK-108277-01), the National Institute for Neurological Disorders and Stroke (R01NS0525851, R21NS072652, R01NS070963, R01NS083534, 5U01NS086625,5U24NS10059103, R01NS105820), and was made possible by the resources provided by Shared Instrumentation Grants 1S10RR023401, 1S10RR019307, and 1S10RR023043. Additional support was provided by the NIH Blueprint for Neuroscience Research (5U01-MH093765), part of the multi-institutional Human Connectome Project. In addition, BF has a financial interest in CorticoMetrics, a company whose medical pursuits focus on brain imaging and measurement technologies. BF's interests were reviewed and are managed by Massachusetts General Hospital and Partners HealthCare in accordance with their conflict of interest policies. 

Data used in the preparation of this article were obtained in part by the OASIS Cross-Sectional with principal investigators D. Marcus, R, Buckner, J, Csernansky J. Morris; P50 AG05681, P01 AG03991, P01 AG026276, R01 AG021910, P20 MH071616, U24 RR021382, and OASIS: Longitudinal: Principal Investigators: D. Marcus, R, Buckner, J. Csernansky, J. Morris; P50 AG05681, P01 AG03991, P01 AG026276, R01 AG021910, P20 MH071616, U24 RR021382. Further, data used in the preparation of this article were obtained from the MIRIAD database. The MIRIAD investigators did not participate in analysis or writing of this report. The MIRIAD dataset is made available through the support of the UK Alzheimer's Society (Grant RF116). The original data collection was funded through an unrestricted educational grant from GlaxoSmithKline (Grant 6GKC). Data collection and sharing for this project was funded by the Alzheimer's Disease Neuroimaging Initiative (ADNI) (National Institutes of Health Grant U01 AG024904) and DOD ADNI (Department of Defense award number  W81XWH-12-2-0012). ADNI is funded by  the National  Institute  on  Aging,  the  National  Institute  of Biomedical  Imaging  and  Bioengineering,  and  through  generous  contributions  from  the  following:  AbbVie, Alzheimer’s Association; Alzheimer’s Drug Discovery Foundation; Araclon Biotech; BioClinica, Inc.; Biogen; Bristol-Myers Squibb Company; CereSpir, Inc.; Cogstate; Eisai Inc.; Elan Pharmaceuticals, Inc.; Eli Lilly and Company; EuroImmun; F. Hoffmann-La Roche Ltd and its affiliated company Genentech, Inc.; Fujirebio; GE Healthcare; IXICO  Ltd.; Janssen Alzheimer Immunotherapy Research  \&  Development, LLC.; Johnson \& Johnson Pharmaceutical  Research \& Development LLC.; Lumosity; Lundbeck; Merck  \&  Co., Inc.; Meso Scale Diagnostics, LLC.; NeuroRx Research;   Neurotrack Technologies; Novartis Pharmaceuticals Corporation; Pfizer Inc.; Piramal Imaging; Servier; Takeda Pharmaceutical Company; and Transition Therapeutics.The Canadian Institutes of Health Research is providing funds to support ADNI clinical sites in Canada. Private sector contributions are facilitated by the Foundation for the National Institutes of Health (www.fnih.org). The grantee  organization is the Northern California Institute for  Research  and  Education, and the study is coordinated by the Alzheimer’s Therapeutic Research Institute at the University of Southern California. ADNI data are disseminated by the  Laboratory for Neuro Imaging at the University of Southern California. Data were also provided in part by the Human Connectome Project, WU-Minn Consortium (Principal Investigators: David Van Essen and Kamil Ugurbil; 1U54MH091657) funded by the 16 NIH Institutes and Centers that support the NIH Blueprint for Neuroscience Research; and by the McDonnell Center for Systems Neuroscience at Washington University.

\section*{References}
\bibliographystyle{elsarticle-num}
\bibliography{mybibfile}
\section*{Appendix}
\setcounter{table}{0}
\renewcommand\thetable{A.\arabic{table}}

\subsection{Datasets}\label{sec:datasets}

\vspace{1ex}
    \textbf{ABIDE II}: The Autism Brain Imaging Data Exchange II ~\cite{abide_dataset} contains cross-sectional data and focuses on autism spectrum disorders covering a wide age rage (5-64 years of age). The dataset contains 1044 MRI scans from 19 different institutions. The 3D magnetization prepared rapid gradient echo (MP-RAGE) sequence, or a vendor specific variant, was used to acquire all data. The corresponding sequence parameters vary depending on the site (see Table 2 of the corresponding paper for details \cite{abide_dataset}). With the exception of a single collection (IP\_1, 1.5 Tesla, Philips Achieva), all MRI data were acquired using 3 Tesla scanners (1 Ingenia and 4 Achieva (Philips), 2 MR750 (GE), 7 TriTim (Siemens), 3 Allegra (Siemens) and 1 Skyra (Siemens)) at voxel resolutions varying from 1.30~mm to 0.7~mm (majority at 1.0~mm) and is available at:\ http://fcon\_1000.projects.nitrc.org/indi/abide/abi\-de\_II.html. 20 cases from the ABIDE-II were used for training.
   
\vspace{1ex}
    \textbf{ADNI}: The Alzheimer’s Disease Neuroimaging Initiative~\cite{ADNI_dataset} was launched in 2003 as a public-private partnership, led by principal investigator Michael W.\ Weiner, MD. The dataset contains 1.5T and 3T-MRI scans acquired at a resolution of 1.0x1.0x1.2~mm with scanners from the three largest MRI vendors (GE, Philips and Siemens) and includes Alzheimer’s disease patients, mild cognitive impaired subjects, and elderly controls. Data were acquired with a MP-RAGE sequence whose parameters are optimized for the different vendors (see \cite{ADNI_mri} for details). The ADNI database has $>$ 2000 participants and is available at:\ http://adni.loni.usc.edu. 40 cases from ADNI where used for training. 180 different cases were used for assessing accuracy and generalizability across disease groups and scanners.
    
\vspace{1ex}
    \textbf{HCP}: The Human Connectome Projects Young Adult~\cite{hcp_dataset} is a cross-sectional dataset and contains 1200 healthy participants from ages 22 to 35. The 3T MR images were acquired with a single scanner (customized Connectome Skyra, Siemens) using a MP-RAGE sequence with a repetition time (TR) of 2.4~s, echo time (TE) of 2.14~ms, an inversion time (TI) of 1~s, and a flip angle of 8$^{\circ}$. Images are 0.7~mm isotropic with a field of view of 224x224, and are de-faced.
    Data is available at:\ https://www.humanconnectome.org/study/hcp-young-adult. 45 cases from HCP were used for assessing accuracy and generalizability to inputs with pre-processing (de-facing, downsampling).
    
\vspace{1ex}
    \textbf{LA5c}: The UCLA Consortium for Neuropsychiatric Phenomics LA5c Study~\cite{LA5c_dataset} is a cross-sectional study and includes 142 participants diagnosed with a neuropsychiatric or neurodevelopmental disorder (schizophrenia, bipolar disorder, ADHD) and 130 normal controls (ages 21-50). T1-weighted MP-RAGE images were acquired on a 3T Siemens Trio at a single\--center with field of view of 250, 256x256 matrix, and 176 1.0~mm sagittal partitions. An TI of 1.1~s, TE of 3.5-3.3~ms, TR of 2.53~s and flip angle of 7$^{\circ}$ was used for all scans. This data was obtained from the OpenfMRI database (https://openfmri.org/dataset/ds000030/). Its accession number is ds000030. 20 cases from LA5c were used for training.
    
\vspace{1ex}
    \textbf{Mindboggle-101}: a manually corrected set of 101 labeled brain images based on a consistent human cortical labeling protocol (DKTatlas) \cite{Klein2012}. It is the largest and most complete set of free, publicly accessible, manually labeled human brain images. The dataset consists of anatomically labeled brain surfaces and volumes derived from T1-weighted brain MRIs of healthy individuals and is available at: \url{https://osf.io/nhtur/}.  Five subjects were scanned specifically for this dataset (MMRR-3T7T-2, Twins-2, and Afterthought-1), all others are from publicly available datasets (i.e.\ Test-Retest OASIS1 \cite{oasis_1_dataset},  the Multi-Modal Reproducibility Resource \cite{Landman2011}, Nathan Kline Institute Test–Retest and Nathan Kline Institute/Rockland Sample, Human Language Network subjects \cite{Morgan2009}, and Colin Holmes 27 template \cite{Holmes1998}, see \cite{Klein2012} for details). The OASIS1 Test-Retest portion (20 subjects) also includes manually labeled subcortical segmentations (labelings by Neuromorphometrics, Inc. (http://Neuromorphometrics.com/) released under Creative Commons License BY-NC-ND 4.0 (http://creativecommons.org/licenses/by-nc-nd/4.0) within Mindboggle-101). Here, we use all volumes with isotropic voxel sizes (78 in total) from Mindboggle-101 to evaluate network performance with respect to a manual reference.
    
\vspace{1ex}
    \textbf{MIRIAD}: The Minimal Interval Resonance Imaging in Alzheimer's Disease~\cite{miriad_dataset} dataset includes longitudinal scans from 46 elderly individuals (ages 55+) diagnosed with Alzheimer's disease and 23 elderly normal controls. The 3D MR images were acquired at a single center with a 1.5T Signa MRI scanner (GE Medical systems), using an IR-FSPGR (inversion recovery prepared fast spoiled gradient recalled) sequence, field of view of 24 cm, 256x256 matrix, 124 1.5~mm coronal partitions (voxel size 0.9x0.9x1.5), TR 15 ms, TE 5.4 ms, flip angle 15$^{\circ}$, and TI 650 ms. The data is available at:\ https://www.ucl.ac.uk/drc/research/methods/min\-imal-interval-resonance-imaging-alzheimers-disease-miriad . 20 cases from MIRIAD were used for validation. 49 different cases were used to assess accuracy and generalizability to a different T1-weighted acquisition sequences (IR-FSGPR) and scanner (GE).

\vspace{1ex}
    \textbf{MMND}: This multi-subject, multi-model neuroimaging dataset was acquired with multiple functional and structural neuroimaging modalities (structural MRI + functional MRI + MEG + EEG) on the same 16 healthy volunteers \cite{MMND}. Here, we only use the structural data which include T1-weighted MPRAGE and Multi-Echo Fast Low Angle Shot (MEF) sequences. The data was collected from a Siemens 3T TIM TRIO with a standard 1~mm isotropic resolution. For each participant, a T1-weighted image was acquired using an MPRAGE sequence (TR 2,250~ms, TE 2.98~ms, TI 900~ms, 190~Hz/pixel; flip angle 9$^{\circ}$) as well as two bandwidth-matched MEF sequences (651~Hz/pixel; TR 20~ms) at both 5$^{\circ}$ and 30$^{\circ}$ flip angles for each of 7 echo-times (TE 1.85~ms; 4.15~ms; 6.45~ms; 8.75~ms; 11.05~ms; 13.35~ms; 15.65~ms). This data was obtained from the OpenNeuro database. Its accession number is ds000117. All MMND cases were used to assess generalizability to MEF sequences.
    
\vspace{1ex}
    \textbf{Oasis-1}~\cite{oasis_1_dataset} and \textbf{Oasis-2}~\cite{oasis_2_dataset}: The Open Access Series of Imaging Studies 1 and 2, both contain scans from a 1.5-T  Vision scanner and a TIM Trio 3T MRI scanner (both Siemens) acquired in a single-center from non-demented and demented individuals diagnosed with very mild to moderate Alzheimer’s disease. All subjects were scanned in sagittal orientation with a voxel resolution of 1.0x1.0x1.25~mm and a MP-RAGE sequence with the following parameters: TR 9.7 ms, TE 4.0 ms, flip angle 10$^{\circ}$, and TI 20 ms. The Oasis-1 set is cross-sectional and contains 416 subject aged 18 to 96. In addition, it contains a test-retest component consisting of 20 subjects that were scanned no more than 90 days apart (all except 5 less than 30 days). The Oasis-2 set focuses on older adults (age 60+) and contains longitudinal scans from 150 subjects. Both dataset are available at:\ https://www.oasis-brains.org/ . 40 cases from Oasis-1 and 20 from Oasis-2 were used for training. 20 different cases from Oasis-1 where used to assess test-retest reliability and 370 cases for quantifying sensitivity to group effects.
    
\vspace{1ex}
    \textbf{THP}: The Traveling Human Phantom~\cite{HTP_dataset} is a dataset collected for assessing multi-site neuroimaging reliability. The THP includes 3D MP-RAGE MRI scans at 1.0~mm isotropic voxel resolution from 5 healthy subjects acquired at 8 different imaging centers. The sites involved in this study had either a Siemens 3T TIM Trio scanner (five sites: IOWA, UMN, UCL, MGH, CCF) or a Philips 3T Achieva scanner (three sites: JHU, DART, UW). The data is available at:\ https://openneuro.org/datasets/ds000206 .  All THP cases were used to quantify accuracy and generalizability across sites and scanners.
 
Participants of the individual studies gave informed consent in accordance with the Institutional Review Board at each of the participating sites. Complete ethic statements are available at the respective study webpages.

\begin{table*}[]
\centering
\caption{Summary of training, validation and testing sets. Table lists the usage, scanner, field strength, (disease) state, age range and number of used subjects for each dataset. }
\begin{tabular}{|l|l|l|l|l|l|l|l|}
\hline
\textbf{Usage}                                        & \textbf{Dataset}               & \textbf{Scanner}     & \textbf{1.5T/3T} & \textbf{State}    & \textbf{Age} & \textbf{Subjects} \\ \hline
\multicolumn{1}{|l|}{\multirow{5}{*}{\textbf{Training}}} & ABIDE-II                       & Phillips             & 3T               & Autism/Normal     & 20-39            & 20              \\
\multicolumn{1}{|c|}{}                                   & ADNI                           & Philips, GE, Siemens & 1.5T/3T          & AD/MCI/Normal     & 56-90                    & 40              \\
\multicolumn{1}{|c|}{}                                   & LA5C                           & Siemens              & 3T               & Neuropsych/Normal & 23-44              & 20              \\
\multicolumn{1}{|c|}{}                                   & OASIS1                         & Siemens              & 1.5T          & Normal            & 18-60              & 40              \\
\multicolumn{1}{|c|}{}                                   & OASIS2                         & Siemens              & 1.5T          & AD/Normal         & 66-90                & 20              \\ \hline
\textbf{Validation}                                      & MIRIAD                         & GE                   & 1.5T            & AD/Normal         & 60-77                & 20              \\ \hline
\multicolumn{1}{|l|}{\multirow{6}{*}{\textbf{Accuracy}}} & ADNI                           & Philips, GE, Siemens & 1.5T/3T          & AD/MCI/Normal     & 58-85                    & 180              \\
\multicolumn{1}{|c|}{}                                   & HCP                            & Siemens              & 3T               & Normal            & 22-35              & 45               \\
\multicolumn{1}{|c|}{}                                   & OASIS1                         & Siemens              & 1.5T          & AD/Normal         & 18-96              & 370               \\
\multicolumn{1}{|c|}{}                                   & MIRIAD                         & GE                   & 1.5 T            & AD/Normal         & 55-80                & 49              \\
\multicolumn{1}{|c|}{}                                   & THP                            & Philips, Siemens     & 3T               & Normal            & -                   & 5                  \\
\multicolumn{1}{|c|}{}                                   & Mindboggle-101                         & Philips, Siemens                   & 1.5T/3T            & Normal         & 19-61                & 78              \\ \hline
\multicolumn{1}{|l|}{\multirow{6}{*}{\textbf{Generalizability}}}                                & ADNI                           & Philips, GE, Siemens & 3T               & AD/MCI/Normal     & 58-85                    & 180                 \\
\multicolumn{1}{|c|}{}                                   & HCP                            & Siemens              & 3T               & Normal            & 22-35              & 45               \\
\multicolumn{1}{|c|}{}                                   & MIRIAD                         & GE                   & 1.5 T            & AD/Normal         & 55-80                & 49              \\
\multicolumn{1}{|c|}{}                                   & MMND                           & Siemens              & 3T               & Normal            & 23-31                   & 16                  \\ 
\multicolumn{1}{|c|}{}                                   & THP                            & Philips, Siemens     & 3T               & Normal            & -                   & 5                \\ 
\multicolumn{1}{|c|}{}                                   & Mindboggle-101                         & Philips, Siemens                   & 1.5T/3T            & Normal         & 19-61                & 78 \\ \hline
\textbf{Reliability}                                     & OASIS1                         & Siemens              & 1.5T          & Normal         & 19-34              & 20                 \\ \hline
\textbf{Sensitivity}                                     & OASIS1                         & Siemens              & 1.5T          & AD/Normal         & 18-96              & 370                   \\ \hline
\end{tabular}
\label{tab:app2}
\end{table*}

\subsection{Labels}
\begin{table*}[]
\centering
\caption{FastSurferCNN (proposed) segmentation labels and mapping to FreeSurfer (FS) for subcortical (left) and cortical (right) structures.}
\begin{tabular}{llll|l|l|l|}
\cline{1-3} \cline{5-7}
\multicolumn{1}{|l|}{\textbf{Subcortical Structures}} & \multicolumn{1}{l|}{\textbf{Proposed}} & \multicolumn{1}{l|}{\textbf{FS}} &  & \textbf{Cortical Structures}      & \textbf{Proposed} & \textbf{FS} \\ \cline{1-3} \cline{5-7} 
\multicolumn{1}{|l|}{Cortical white matter (lh)}      & \multicolumn{1}{l|}{1}                 & \multicolumn{1}{l|}{2}                   &  & caudalanteriorcingulate (lh)      & 34                & 1002                \\
\multicolumn{1}{|l|}{Lateral Ventricle (lh)}          & \multicolumn{1}{l|}{2}                 & \multicolumn{1}{l|}{4}                   &  & caudalmiddlefrontal (lh, rh)      & 35                & 1003, 2003          \\
\multicolumn{1}{|l|}{Inferior Lateral Ventricle (lh)} & \multicolumn{1}{l|}{3}                 & \multicolumn{1}{l|}{5}                   &  & cuneus (lh)                       & 36                & 1005                \\
\multicolumn{1}{|l|}{Cerebellar White Matter (lh)}    & \multicolumn{1}{l|}{4}                 & \multicolumn{1}{l|}{7}                   &  & entorhinal (lh, rh)               & 37                & 1006, 2006          \\
\multicolumn{1}{|l|}{Cerebellar Cortex (lh)}          & \multicolumn{1}{l|}{5}                 & \multicolumn{1}{l|}{8}                   &  & fusiform (lh, rh)                 & 38                & 1007, 2007          \\
\multicolumn{1}{|l|}{Thalamus (lh)}                   & \multicolumn{1}{l|}{6}                 & \multicolumn{1}{l|}{10}                  &  & inferiorparietal (lh, rh)         & 39                & 1008, 2008          \\
\multicolumn{1}{|l|}{Caudate (lh)}                    & \multicolumn{1}{l|}{7}                 & \multicolumn{1}{l|}{11}                  &  & inferiortemporal (lh, rh)         & 40                & 1009, 2009          \\
\multicolumn{1}{|l|}{Putamen (lh)}                    & \multicolumn{1}{l|}{8}                 & \multicolumn{1}{l|}{12}                  &  & isthmuscingulate (lh)             & 41                & 1010                \\
\multicolumn{1}{|l|}{Pallidum (lh)}                   & \multicolumn{1}{l|}{9}                 & \multicolumn{1}{l|}{13}                  &  & lateraloccipital (lh, rh)         & 42                & 1011, 2011          \\
\multicolumn{1}{|l|}{3rd-Ventricle}                   & \multicolumn{1}{l|}{10}                & \multicolumn{1}{l|}{14}                  &  & lateralorbitofrontal (lh)         & 43                & 1012                \\
\multicolumn{1}{|l|}{4th-Ventricle}                   & \multicolumn{1}{l|}{11}                & \multicolumn{1}{l|}{15}                  &  & lingual (lh)                      & 44                & 1013                \\
\multicolumn{1}{|l|}{Brain Stem}                      & \multicolumn{1}{l|}{12}                & \multicolumn{1}{l|}{16}                  &  & medialorbitofrontal (lh)          & 45                & 1014                \\
\multicolumn{1}{|l|}{Hippocampus (lh)}                & \multicolumn{1}{l|}{13}                & \multicolumn{1}{l|}{17}                  &  & middletemporal (lh, rh)           & 46                & 1015, 2015          \\
\multicolumn{1}{|l|}{Amygdala (lh)}                   & \multicolumn{1}{l|}{14}                & \multicolumn{1}{l|}{18}                  &  & parahippocampal (lh)              & 47                & 1016                \\
\multicolumn{1}{|l|}{CSF}                             & \multicolumn{1}{l|}{15}                & \multicolumn{1}{l|}{24}                  &  & paracentral (lh)                  & 48                & 1017                \\
\multicolumn{1}{|l|}{Accumbens (lh)}                  & \multicolumn{1}{l|}{16}                & \multicolumn{1}{l|}{26}                  &  & parsopercularis (lh, rh)          & 49                & 1018, 2018          \\
\multicolumn{1}{|l|}{Ventral DC (lh)}                 & \multicolumn{1}{l|}{17}                & \multicolumn{1}{l|}{28}                  &  & parsorbitalis (lh, rh)            & 50                & 1019, 2019          \\
\multicolumn{1}{|l|}{Choroid Plexus (lh)}             & \multicolumn{1}{l|}{18}                & \multicolumn{1}{l|}{31}                  &  & parstriangularis (lh, rh)         & 51                & 1020, 2020          \\
\multicolumn{1}{|l|}{Cortical white matter (rh)}      & \multicolumn{1}{l|}{19}                & \multicolumn{1}{l|}{41}                  &  & pericalcarine (lh)                & 52                & 1021                \\
\multicolumn{1}{|l|}{Lateral Ventricle (rh)}          & \multicolumn{1}{l|}{20}                & \multicolumn{1}{l|}{43}                  &  & postcentral (lh)                  & 53                & 1022                \\
\multicolumn{1}{|l|}{Inferior Lateral Ventricle (rh)} & \multicolumn{1}{l|}{21}                & \multicolumn{1}{l|}{44}                  &  & posteriorcingulate (lh)           & 54                & 1023                \\
\multicolumn{1}{|l|}{Cerebellar White Matter (rh)}    & \multicolumn{1}{l|}{22}                & \multicolumn{1}{l|}{46}                  &  & precentral (lh)                   & 55                & 1024                \\
\multicolumn{1}{|l|}{Cerebellar Cortex (rh)}          & \multicolumn{1}{l|}{23}                & \multicolumn{1}{l|}{47}                  &  & precuneus (lh)                    & 56                & 1025                \\
\multicolumn{1}{|l|}{Thalamus (rh)}                   & \multicolumn{1}{l|}{24}                & \multicolumn{1}{l|}{49}                  &  & rostralanteriorcingulate (lh, rh) & 57                & 1026, 2026          \\
\multicolumn{1}{|l|}{Caudate (rh)}                    & \multicolumn{1}{l|}{25}                & \multicolumn{1}{l|}{50}                  &  & rostralmiddlefrontal (lh, rh)     & 58                & 1027, 2027          \\
\multicolumn{1}{|l|}{Putamen (rh)}                    & \multicolumn{1}{l|}{26}                & \multicolumn{1}{l|}{51}                  &  & superiorfrontal (lh)              & 59                & 1028                \\
\multicolumn{1}{|l|}{Pallidum (rh)}                   & \multicolumn{1}{l|}{27}                & \multicolumn{1}{l|}{52}                  &  & superiorparietal (lh, rh)         & 60                & 1029, 2029          \\
\multicolumn{1}{|l|}{Hippocampus (rh)}                & \multicolumn{1}{l|}{28}                & \multicolumn{1}{l|}{53}                  &  & superiortemporal (lh, rh)         & 61                & 1030, 2030          \\
\multicolumn{1}{|l|}{Amygdala (rh)}                   & \multicolumn{1}{l|}{29}                & \multicolumn{1}{l|}{54}                  &  & supramarginal (lh, rh)            & 62                & 1031, 2031          \\
\multicolumn{1}{|l|}{Accumbens (rh)}                  & \multicolumn{1}{l|}{30}                & \multicolumn{1}{l|}{58}                  &  & transversetemporal (lh, rh)       & 63                & 1034, 2034          \\
\multicolumn{1}{|l|}{Ventral DC (rh)}                 & \multicolumn{1}{l|}{31}                & \multicolumn{1}{l|}{60}                  &  & insula (lh, rh)                   & 64                & 1035, 2035          \\
\multicolumn{1}{|l|}{Choroid Plexus (rh)}             & \multicolumn{1}{l|}{32}                & \multicolumn{1}{l|}{63}                  &  & caudalanteriorcingulate (rh)      & 65                & 2002                \\
\multicolumn{1}{|l|}{WM-hypointensities}              & \multicolumn{1}{l|}{33}                & \multicolumn{1}{l|}{77}                  &  & cuneus (rh)                       & 66                & 2005                \\ \cline{1-3}
                                                      &                                        &                                          &  & isthmuscingulate (rh)             & 67                & 2010                \\
                                                      &                                        &                                          &  & lateralorbitofrontal (rh)         & 68                & 2012                \\
                                                      &                                        &                                          &  & lingual (rh)                      & 69                & 2013                \\
                                                      &                                        &                                          &  & medialorbitofrontal (rh)          & 70                & 2014                \\
                                                      &                                        &                                          &  & parahippocampal (rh)              & 71                & 2016                \\
                                                      &                                        &                                          &  & paracentral (rh)                  & 72                & 2017                \\
                                                      &                                        &                                          &  & pericalcarine (rh)                & 73                & 2021                \\
                                                      &                                        &                                          &  & postcentral (rh)                  & 74                & 2022                \\
                                                      &                                        &                                          &  & posteriorcingulate (rh)           & 75                & 2023                \\
                                                      &                                        &                                          &  & precentral (rh)                   & 76                & 2024                \\
                                                      &                                        &                                          &  & precuneus (rh)                    & 77                & 2025                \\
                                                      &                                        &                                          &  & superiorfrontal (rh)              & 78                & 2028                \\ \cline{5-7} 
\end{tabular}
\label{tab:labels}
\end{table*}

\end{document}